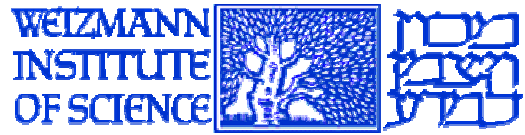

# Broadband Spectrally Correlated Light - Properties, Sources and Applications

By

Avi Pe'er

Under the Supervision of Prof. Asher A. Friesem

Submitted to the Scientific Council of the Weizmann Institute of Science in partial fulfillment of the requirements for the degree of

Doctor of Philosophy

At the

Weizmann Institute of Science

Rehovot, Israel

May 2005

# Abstract

The coherence properties of broadband down converted light are investigated theoretically and experimentally in both the classical and quantum mechanical frameworks. Although broadband down converted light is one-photon incoherent noise, it possesses unique two-photon coherence. Thus, for performing two-photon processes, down-converted light is equivalent to coherent ultrashort pulses in some aspects and to coherent CW laser radiation in others. This equivalence is theoretically established and experimentally demonstrated in the classical, high-power regime. Two applications that take advantage of multi-photon coherence are then described in detail. One exploits the unique two-photon coherence of broadband down converted light for optical spread spectrum communication. Another demonstrates the ability to beat the diffraction limit in optical lithography by a factor of $N$ with classical light that possesses $N$-photon coherence, either pulsed or down-converted light. A key part of this research is the design and implementation of an efficient source that generates high-power broadband down conversion in a cavity. A broadband oscillation in the cavity is ensured by a mechanism very similar to passive mode-locking of ultrashort pulse laser sources, which we term as pair-wise mode locking, thus extending the equivalence to ultrashort pulses to the light generating sources also. Finally, the quantum mechanical properties of broadband down-conversion are explored experimentally, by performing a two-photon process (sum frequency generation) at the low power level of entangled photon pairs. The non-classical nature of the light is expressed by a linear intensity dependence of the non-linear process. Temporal shaping with femtosecond resolution of the entangled photon pairs is demonstrated using this non-linear interaction.

I certify that I have read this dissertation and that in my opinion it is fully adequate, in scope and quality, as a dissertation for the degree of Doctor of Philosophy.

Professor Asher A. Friesem ………………………….
Advisor

# Acknowledgments

First, I wish to express my deep gratitude to my adviser Prof. Asher Friesem. Apart from being an excellent scientist, Asher was for me an ideal advisor. I found in him a unique openness to new ideas, no matter how crazy, combined with a calm approach and a sense of humor. I believe that the wonderful working atmosphere in the group is greatly inspired by him.

I owe a large debt to Prof. Yaron Silberberg, my "virtual" adviser. Yaron's deep questions were the drive for major progress in my work and his insight and experience were a great source for advice, scientific or other. I also thank Prof. Nir Davidson for advice on numerous occasions. Nir's vast experience and deep understanding were a great help in the difficult moments of experiments. Already in the early (risky) stages, Nir showed confidence in my research and encouraged me to pursue an academic career, and for that I am especially grateful.

Special thanks are due to Barak Dayan, my partner and friend. Barak's persistence and strive for perfection in either experiment or theoretical understanding, were a vital ingredient in the quality of our results. We complemented each other in so many ways that the term "entangled pair" seems adequate, both in science and in life. I thank my colleagues, the students in the optics group, for many fruitful discussions, for help with equipment and for the great time I had working in their company. Specifically, I thank Nirit Dudovich for her friendship, wisdom and humble nature.

I wish to thank Prof. Amnon Yogev and Prof. Daniel Zajfman, for allowing us to conduct our experiments in their labs and to use their laser sources. Many thanks to the technical support team of the department of complex systems for their important part in the realization of my experiments: to Uri for "crazy" electronics, to Yossi Shopen and his team for creative opto-mechanics, to Yossi Drier for being so efficient with our computers and to Rosti for help with almost any technical need. I thank our administrative staff – Israel, Rachel, Perla and Malka, for handling the really important stuff for us and for easing our life in the department more than one can hope for.

I thank with pride my parents, Rivka and Eytan, for so much of what I am today. The fruits I now harvest are very much due to their education and belief in my abilities. I heartedly thank my wife's parents, Shoshana and Shimon Lev, for their invaluable help that is unconditionally given whenever needed.

Finally, I wish to thank my wife Idit, who apart from supporting and loving me so much was also a secret partner in this research. Her valuable experimental advice was never appropriately acknowledged. My last thanks go to Ouri and Dan, my two wonderful children, the true light in my life.

*To Mama and Papa,
my grandparents*

# TABLE OF CONTENTS

# Chapter 1
# Introduction

Parametric processes in non-linear crystals, such as parametric down-conversion, sum frequency generation (parametric up-conversion) and second harmonic generation were widely investigated over the last 30 years and many devices that exploit these processes, such as optical parametric oscillators (OPO) and optical parametric amplifiers (OPA), were built. Theoretical treatment of these processes was given in both the classical and quantum mechanical frameworks.[1-14]

Quantum mechanically, the parametric down conversion process, schematically shown in figure 1, can be viewed as the conversion of one energy quantum at a high frequency (a pump photon) via the mediation of a non-linear crystal, into two lower energy quanta (signal photon and idler photon). figure 2 shows the process of sum frequency generation - the symmetric inverse process. Both processes can either be spontaneous or stimulated. Spontaneous parametric down conversion is one of the key methods for generation of entangled photon pairs in experiments on quantum optics.[15-20] We focus our discussion on time-energy entangled pairs that are generated when a pump photon of a well-defined energy (frequency) is spontaneously broken into two, a signal and an idler photon, whose energy spectrum is broad. Such a pair of photons has interesting properties. One property results from the fact that although the energy spectrum of each photon is broad, limited only by the phase matching properties of the non-linear crystal, the sum of their energies must be equal to the pump energy, which is very well defined. Thus, by measuring the energy of one photon it is possible to deduce the energy of the other, even though its energy was never measured. Another important property of entangled photon pairs is that although these two photons are CW photons, i.e. the probability to detect one of them extends over a long time (the coherence time of the pump), they arrive at a detector exactly at the same time, with the temporal accuracy one expects from two transform limited pulses of the same spectral width. This accuracy, which may be as short as tens of femtoseconds, has a peculiar consequence – we don't know when each of the photons arrives, but we do know they arrive together. Thus, performing a sum frequency generation experiment (or two photon absorption) with an entangled pair[21] [21-24] enables a spectral resolution equivalent to what is expected when using single frequency photons (CW) together with the temporal resolution expected from an experiment with transform-limited pulses.

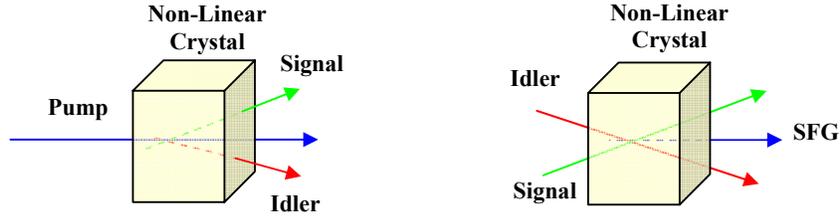

Figure 1: Parametric Down Conversion   Figure 2: Sum Frequency Generation

In a cavity (OPO) or when the pump intensity is very high (OPA), the parametric down conversion becomes stimulated, generating a macroscopic number of down converted photons, which can be described by the classical equations of non-linear optics. Although the field can be considered "classical" and the quantum description is no longer entangled as in the case of single photon pairs, correlations still exist between the complex amplitudes of the signal and idler fields. Specifically, although each field in itself is incoherent white noise, the amplitudes of twin frequencies are complex conjugates of each other, as

$$A_s(\omega) = A_i^*(\omega_p - \omega), \qquad (1)$$

where the subscripts *s,i* and *p* denote the signal, idler and pump respectively [3,25]. This symmetry in the spectrum indicates that the temporal envelope of the down-converted field $A(t)$ around the center frequency $\omega_p/2$ is real; i.e. it has only one quadrature in the complex plane [7-9,12,26]. Accordingly, broadband down-converted light is indeed incoherent noise, yet it's phase can only take the values of zero or π (randomly). The symmetry can be easily derived from classical analysis. However, since in the down conversion process only pairs of photons are generated, the symmetry is satisfied also quantum mechanically to the last photon without any noise, assuming an ideal lossless situation. The quantum mechanical squeezing properties of the light are just a consequence of this exact symmetry.

The special properties of two-photon processes, such as two-photon absorption (TPA) and sum frequency generation (SFG), when performed with down converted light, are understood in the context of the exact symmetry in the spectrum of the light. In general, the TPA probability (or SFG intensity) at frequency $\Omega$ is given by

$$R(\Omega) \propto \left| \int d\omega A_s(\omega) A_i(\Omega - \omega) \right|^2. \tag{2}$$

When $\Omega \neq \omega_p$, the integrand contains uncorrelated signal-idler frequency pairs, leading to a negligible two-photon excitation, as expected from two-photon processes with incoherent white noise. Yet, when $\Omega = \omega_p$, the random phase of the integrand is cancelled out, since the summation is over correlated frequency pairs, leading to a fully constructive interference, of

$$R(\Omega) \propto \left| \int d\omega |A_s(\omega)|^2 \right|^2. \tag{3}$$

This result is exactly equivalent to that obtained when the process is performed with coherent ultrashort pulses; however it is valid only within the two-photon spectrum of the pump laser, which can be arbitrarily narrow. This seemingly contradicting coexistence of an ultrashort temporal resolution along with an arbitrarily fine spectral resolution forms the backbone of my Ph.D research.

It is well known that an ultrashort pulse is just a private case in a broad family of field solutions that can efficiently excite narrowband two-photon transitions. In fact, any field with just one quadrature is equally suitable, as was demonstrated in the past by shaping of coherent ultrashort pulses [27-29]. In a sense, my research focuses on the extreme case where the spectral phase of the field is completely random and incoherent, yet the required symmetry is guaranteed by the physical process [30,31].

Just like with ultrashort pulses, TPA with down-converted light is efficient due to the constructive interference of all the possible frequency pairs. As a result, even though the down-converted light is neither coherent nor pulsed, the process can be controlled just as well by coherent pulse shaping techniques. Since the phases of the single frequency components of broadband down-converted light are undetermined, one can define the light as *one-photon incoherent*. However, down-converted light is still *two-photon coherent* (all the frequency pairs share the same phase) and this broadband two-photon coherence is the one that allows the process to be coherently shaped and controlled. Specifically, the TPA (or SFG) amplitude can be reversibly eliminated by a minute relative delay / dispersion between the signal and the idler beams [25,30-32]. This feature stands in the heart of all the applications that will be discussed in the following chapters.

A corner stone of my Ph.D research is the development of an efficient low-threshold OPO source that emits broadband down-converted light. Since mode competition in an oscillator cavity normally causes significant narrowing of the emitted spectrum, the design of such an OPO cavity is non-trivial and must incorporate some mechanism to suppress mode competition. The unique coherence properties and the equivalence to ultrashort pulses are exploited to the extreme in this cavity design, and the similarity of the OPO cavity configuration to that of mode-locked ultrafast lasers lead us to coin the term "pair-wise mode locking" to describe this technique.

Most of my research was conducted in the high power regime, where most of the properties of down-converted light are adequately described by classical equations. Yet, since the physical process generates pairs of photons, the signal and the idler fields are ideally correlated to the last photon without any noise. Although such non-classical correlation exists at high powers, our experiments could not resolve it. However, at the low power regime of single photon pairs, the correlation becomes crucial and we conducted several experiments to demonstrate and explore it. Specifically, we performed a two-photon nonlinear interaction (sum-frequency generation) with entangled photons, where the non-classical signature is a linear flux dependence of the non-linear process. We also exploited this interaction between the photons as a coincidence detector to measure the temporal correlation between the photons and shape it with a femtosecond resolution.

The organization of the rest of this thesis is as follows: Chapter 2 describes the verification of the special two-photon coherence properties of down-converted light in a TPA experiment in atomic Rubidium. Two possible applications are discussed in chapters 3 and 4 – one in optical communication and the other in lithography. Chapters 5 and 6 are devoted to the pair-wise mode locked OPO source. Chapter 5 describes my experimental work and chapter 6 includes a theoretical performance analysis of the OPO source, so far unpublished. Chapter 7 describes the experiments at the low-power entangled photons regime and my conclusions and future outlook are summarized in chapter 8

# Chapter 2
# Two Photon Absorption and Coherent Control with Broadband Down-Converted Light

To my knowledge, only one experimental investigation was devoted in the past to the special coherence properties of broadband down-converted light [31], where relatively sharp temporal and spectral resolutions were demonstrated with SFG.

In this work we demonstrated the equivalence between incoherent down-converted light and shaped coherent ultrashort pulses. By performing a TPA experiment in atomic Rb, we were able to demonstrate both the sharp spectral resolution and the ultrafast temporal resolution. We measured the sharp dependence of the TPA rate on the relative delay between the signal and the idler fields and then coherently shaped this dependence using pulse shaping techniques.

Our experiments establish the important two-photon coherence properties of broadband down-converted light and serve as the basis for all the work described in the following chapters, either on high power sources and applications or on low power non-classical properties.

# Two Photon Absorption and Coherent Control with Broadband Down-Converted Light

Barak Dayan, Avi Pe'er, Asher A. Friesem, and Yaron Silberberg

*Department of Physics of Complex Systems, Weizmann Institute of Science, Rehovot 76100, Israel*

We experimentally demonstrate two-photon absorption with broadband down-converted light (squeezed vacuum). Although incoherent and exhibiting the statistics of a thermal noise, broadband down-converted light can induce two-photon absorption with the same sharp temporal behavior as femtosecond pulses, while exhibiting the high spectral resolution of the narrow band pump laser. Using pulse-shaping methods, we coherently control two-photon absorption in rubidium, demonstrating spectral and temporal resolutions that are 3–5 orders of magnitude below the actual bandwidth and temporal duration of the light itself. Such properties can be exploited in various applications such as spread-spectrum optical communications, tomography, and nonlinear microscopy.

In two-photon absorption (TPA), two photons whose sum energy equals that of an atomic transition must arrive at the atom together. The two, seemingly contradicting, demands of a narrow temporal and a narrow spectral behavior of the inducing light are typically maximized by transform-limited pulses, which exhibit the highest peak intensity possible for a given spectral bandwidth. Nonetheless, it was shown [1,2] that pulses can be shaped in a way that will stretch them temporally yet will not affect the transition probability, and even increase it in certain cases [3]. Other experiments have exploited coherent control [4–7] to increase the spectral selectivity of nonlinear interactions induced by ultrashort pulses [8–10]; however, the spectral resolution demonstrated by these methods remains considerably inferior to that obtained by narrow band, continuous lasers. A few experiments [11,12] have performed TPA with coherent, narrow band down-converted light, demonstrating nonclassical features which appear at very low powers [13–15] and result from the time and energy correlations (entanglement) between the down-converted photon pairs [16–18]. At high power levels (as those discussed in this work) these correlations vanish, yet similarly nonclassical phase and amplitude correlations [19,20] appear between the signal and idler beams. Unlike the time and energy correlations at low powers, these phase and amplitude correlations cannot be described in the usual form of second-order coherence. At sufficiently high powers, which greatly exceed the single-photons regime, broadband down-converted light that is pumped by a narrow band laser is inherently incoherent and exhibits the properties of a broadband thermal noise [16,21]; consequently, it is not expected to be effective at inducing TPA. Nevertheless, it was shown that at the appropriate conditions, the quantum correlations within the down-converted spectrum can give rise to efficient sum-frequency generation [22,23].

In this work we show that down-converted light beams with a spectral bandwidth that significantly exceeds that of the pump can induce TPA just like transform-limited pulses of the same bandwidth. Consequently, the interaction exhibits a sharp, ultrashort pulse-like temporal behavior and can be coherently controlled by pulse-shaping techniques, even though the down-converted light beams are neither coherent nor pulsed [24]. This effect occurs as long as the transition energy lies within the spectrum of the pump laser that generated the light; thus, the spectral selectivity of the interaction is dictated by the narrow band pump laser and not by the orders of magnitude wider bandwidth of the down-converted light itself. We demonstrate these principles experimentally by inducing and coherently controlling TPA in atomic rubidium with down-converted light and by obtaining results that are practically identical to those obtained with coherent ultrashort pulses.

The underlying principle that enables coherent TPA with broadband down-converted light is based on the fact that the quantum interference that governs TPA involves pairs of electromagnetic modes. Since the excitation of an atomic level with frequency $\Omega$ may be induced by any two photons with frequencies $\omega$ and $\Omega - \omega$, regardless of the exact value of $\omega$, the final population $p_f$ is proportional to [1]

$$p_f \propto \left| \int E(\omega) E(\Omega - \omega) d\omega \right|^2, \quad (1)$$

where $E(\omega)$ is the spectral amplitude of the light. As is obvious from Eq. (1), the individual phases of the spectral modes do not matter; only the phase of the product $E(\omega)E(\Omega - \omega)$ is important. Despite the incoherence of each of the down-converted signal and idler beams, they exhibit exactly this mutually coherent phase behavior at frequency pairs, due to the inherent phase and amplitude quantum correlations within the down-converted

spectrum [19]:

$$\frac{1}{\sqrt{\omega}} E_s(\omega) \approx \frac{1}{\sqrt{\omega_p - \omega}} E_i^*(\omega_p - \omega), \quad (2)$$

where $\omega_p$ is the pump frequency and $E_s(\omega)$, $E_i(\omega)$ denote the spectral amplitudes of the signal and the idler beams, respectively. In the case of TPA induced by the combined signal and idler beams from the same source, these correlations have a drastic effect when the pump frequency is equal to the total transition frequency. Combining Eqs. (1) and (2), letting $\omega_p = \Omega$, reveals that the random phase of $E_s(\omega)$ is always compensated by the opposite phase of $E_i(\omega_p - \omega)$. Thus, the integrand in Eq. (1) has a constant phase, leading to a full constructive interference of all the spectral combinations, exactly as if the interaction was induced by a pair of transform-limited pulses with the same spectra as the signal and idler beams. Moreover, this TPA process will be sensitive to minute delays between the signal and idler beams, to dispersion, and even to pulse shaping, exactly as if it was induced by a pair of ultrashort pulses (this principle holds for sum-frequency generation as well, when the final wavelength is equal to that of the pump [23]). Since this constructive interference occurs only when the final state energy falls within the spectrum of the pump laser, the spectral resolution of the TPA process will be equal to the spectral bandwidth of the pump laser, regardless of the actual bandwidth of the down-converted light itself. Note that in the case of a continuous pump laser, this implies a possible spectral resolution of a few MHz or even less, which is phenomenally high for an interaction that is induced by light with a spectral bandwidth that may be 5–7 orders of magnitude broader.

A complete quantum-mechanical analytic calculation [25], which takes into account the spectral bandwidths of the pump laser and of the atomic level, shows that the TPA signal is composed of two parts, which may be referred to as the "coherent TPA" and the "incoherent TPA." The coherent TPA results from the summation of conjugated spectral components and indeed can be coherently controlled. The incoherent TPA, however, results from the summation of all other random spectral combinations and is a direct result of the incoherence of the down-converted light; therefore, it is unaffected by spectral-phase manipulations and may be regarded as a background noise, which limits the equivalence of the down-converted light to a coherent pulse. The ratio between the coherent term $I^c$ and the incoherent term $I^{ic}$ can be approximated by

$$\frac{I^c}{I^{ic}} \approx \frac{B}{(\gamma_p + \gamma_f)} \frac{n^2 + n}{n^2}, \quad (3)$$

where $n$ is the spectral average of the mean photon flux, and $B$, $\gamma_p$, $\gamma_f$ are the bandwidths of the down-converted light, the pump laser, and the final state, respectively. This expression reveals the importance of using spectrally broad down-converted light. The coherent term becomes dominant only when the down-converted bandwidth exceeds both the pump bandwidth and the final level width: $B > (\gamma_p + \gamma_f)(\frac{n^2}{n^2+n})$. Equation (3) shows that the coherent term exhibits a linear intensity dependence at low powers, as was previously predicted [13,14] and experimentally observed [11].

Our experimental layout is described in Fig. 1. The pump laser (Spectra-Physics MOPO-SL laser) was tuned to emit 3 ns pulses with a bandwidth of 0.04 nm around 516.65 nm, which corresponds to the $5S - 4D$ transition in Rb [Fig. 1(a)]. These pulses pumped a BBO crystal that was located inside a low-finesse resonator, with the phase-matching conditions chosen to obtain broadband

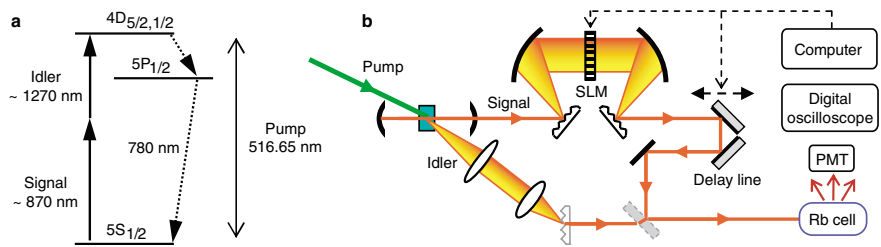

FIG. 1 (color online). (a) Atomic energy levels of Rb, showing the induced two-photon transition and the resulting cascaded decay. (b) The experimental layout; 3 ns pulses at 516.65 nm were used to pump a 14 mm long beta barium borate (BBO) crystal in a low-finesse resonator. The relative angles between the crystal axis, the pump laser, and the resonator were chosen to obtain broadband phase matching for noncollinear type-I down-conversion. Unlike the signal beam, whose direction of propagation was set by the resonator, the idler beam had a wavelength-dependent angular spread, due to the phase matching conditions in the crystal. Therefore, it was directed through an imaging system that included a transmission grating which compensated for this spread. The signal beam was directed through a pulse shaper and a computer-controlled delay line with 0.1 $\mu$ resolution and then combined with the idler beam by a dichroic mirror. The pulse shaper separates the spectral components of the beam and utilizes a computer-controlled spatial light modulator (SLM) to introduce the desired spectral phase filter to the light. A photomultiplier tube (PMT) and a digital oscilloscope were used to perform a triggered measurement of the TPA-induced fluorescence.

(~100 nm each), noncollinear signal and idler beams. The signal beam was directed through a computer-controlled pulse shaper, which is normally used to temporally shape femtosecond pulses and performs as a spectral-phase filter [26]. From the pulse shaper the signal beam continued to a computer-controlled delay line and then was combined with the idler beam by a dichroic mirror. The combined beams entered the Rb cell, and the induced TPA was measured through the resulting fluorescence at 780 nm. Our analysis predicts that the down-converted beams, each 3 ns long with a peak power of about 1 MW and a spectral bandwidth of about 100 nm, should induce TPA with the same efficiency and temporal resolution as 23 fs pulses with a peak power of about 150 GW, while exhibiting a spectral resolution of 0.04 nm.

To verify these predictions experimentally, we scanned both the pump wavelength and the relative delay between the signal and idler beams. Figure 2 shows the TPA signal versus the signal-idler relative delay and the pump wavelength. As is evident, for off-resonance pump or for large signal-idler delays, only low efficiency TPA signal is observed, which is insensitive to either pump wavelength or signal-idler delay. This signal corresponds to the classical, incoherent TPA background signal. However, when the pump wavelength was set to the $5S - 4D$ transition resonance, an additional, high efficiency, TPA signal appeared. The sharp response of this coherent TPA signal to a few femtoseconds delay between the beams is practically identical to the case of TPA induced by two coherent, 23 fs pulses. This temporal resolution is 5 orders of magnitude better than the actual 3 ns temporal duration of the down-converted light.

Since the peak power of the equivalent transform-limited pulses is extremely high (150 GW), the signal and idler beams had to be expanded and attenuated in order to avoid complete saturation of the transition. Unfortunately, we could not attenuate the beams enough to avoid saturation completely, due to high levels of noise in our system. As a result, our measurements had to be performed in a partially saturated regime, where the measured intensity dependence of the TPA process was less than quadratic, and the $4D$ level was power broadened. Thus, the observed 0.12 nm spectral width of the coherent TPA is dictated by the bandwidth of the pump laser (0.04 nm) and the width of the (power-broadened) $4D$ level (~0.08 nm). This spectral resolution is 2000 times narrower than the total bandwidth of the down-converted light.

Finally, we demonstrated quantum coherent control over the coherent TPA process, in a similar way to coherent control of TPA with ultrashort pulses [1]. For that we used the pulse shaper to apply a square-wave phase filter on the signal spectrum [Fig. 3(a)]. With ultrashort pulses this has the effect of splitting the pulse temporally to a train of several smaller pulses. Figure 3(b) shows precisely this behavior of the coherent TPA signal as a function of the signal-idler delay. Figure 3(c) depicts the experimental and theoretical TPA signal at zero delay as a function of the magnitude of the square wave phase filter. The results, which are identical to those obtained in coherent control experiments with femtosecond pulses, show the cyclic transition between complete constructive and destructive quantum interference of the different spectral sections and demonstrate the ability to fully control the coherent TPA process.

Broadband down-conversion of a narrow band pump can be considered as an optical spread-spectrum source, which generates both a broadband white noise key and its conjugate simultaneously [27]. A spread-spectrum communication channel can therefore be established by modulating the phase of one of the keys at the transmitter and using TPA or sum-frequency generation to reveal the resulting modulations of the coherent signal at the receiver. Note that any phase modulations performed on the incoherent signal beam (or even the mere presence of the signal beam, in the existence of background noise) cannot be detected without the idler beam. Furthermore, many such communication channels can share the same signal and idler beams by assigning a unique signal-idler delay for each channel, thus creating an optical code division

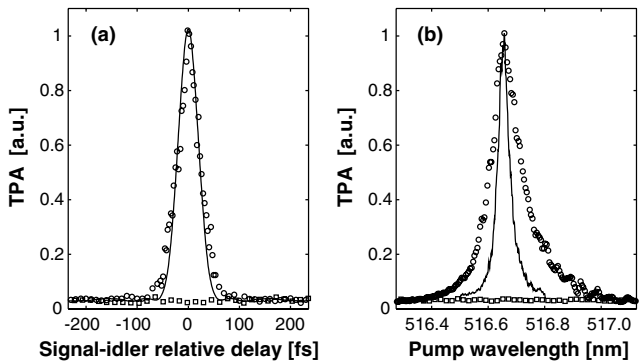

FIG. 2.  Experimental TPA with broadband down-converted light, as a function of the signal-idler delay and the pump wavelength. The graphs clearly show a coherent TPA signal that is narrow both temporally and spectrally, superimposed on an incoherent background signal, which is insensitive to both pump wavelength and signal-idler relative delay. In this experiment the pulse shaper was used only to compensate for the dispersion in our system. (a) Calculated (line) and experimental TPA with off-resonance (squares) and on-resonance (circles) pump, as a function of the signal-idler delay. (b) Experimental TPA at zero signal-idler delay (circles), and at 100 fs signal-idler delay (squares), as a function of the pump center wavelength, together with a typical spectrum of the pump (line). The coherent TPA signal appears only when the pump is on-resonance with the $5S - 4D$ transition, and exhibits a sharp dependence on the signal-idler delay, exactly as if the interaction was induced by a pair of coherent, 23 fs pulses with the same spectra as the signal and idler beams.

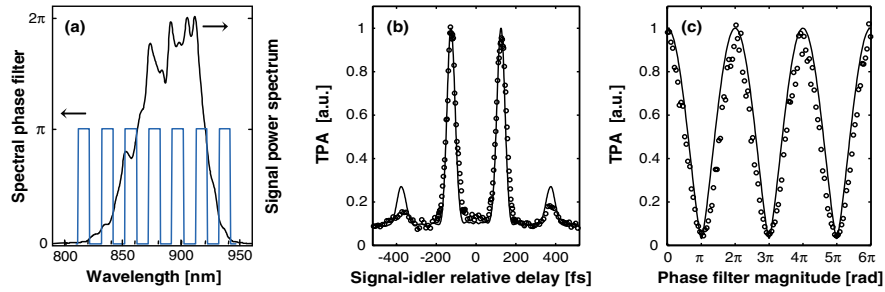

FIG. 3 (color online). Quantum coherent control of TPA with broadband down-converted light. (a) The square-wave spectral phase filter applied on the signal beam, together with a typical power spectrum of the signal. (b) Experimental (circles) and calculated (line) TPA signal as a function of the signal-idler delay, using the spectral phase filter described in (a). The splitting of the central peak confirms that the coherent TPA behaves exactly as if it was induced by a pair of transform-limited pulses, with one of them temporally shaped by the applied phase filter. (c) Experimental (circles) and calculated (line) TPA signal as a function of the magnitude of the square wave spectral phase filter. The graph shows the periodic annihilation of the coherent TPA at magnitudes of $\pi, 3\pi, 5\pi$, where a complete destructive quantum interference occurs ("dark pulse"), and the complete reconstruction of the TPA signal at amplitudes of $2\pi, 4\pi, 6\pi$, etc.

multiple access network [27]. Other applications may include nonlinear microscopy and tomography, where the high efficiency and spatial resolution of ultrashort pulses can be obtained with continuous, nondamaging intensities.

We wish to thank Daniel Zajfman for the loan of the laser system, Adi Diner for advice and assistance, and Nir Davidson for many helpful discussions.

# Chapter 3
# Optical Code-Division Multiple Access using Broadband Parametrically Generated Light

The unique two-photon coherence of broadband down-converted light is appealing for several applications. We investigated one major application - optical code division multiple access (CDMA), which is an advanced scheme for multiplexing many user channels onto a single communication channel based on the method of direct sequence spread spectrum[33]. In CDMA, all user channels utilize all the available bandwidth all the time. Interference is prevented by assigning a unique broadband key that ideally resembles white noise to each user. The advantages of CDMA are well recognized, making it also very attractive for optical communication. The main obstacle for optical CDMA is how to generate the broadband key, since the required bandwidth is much broader than achievable electronically[34-40]. In parametric down-conversion the signal and idler are two broadband white noise fields that are complex conjugates of each other. This correlation makes these fields an ideal set of key and conjugate key. Accordingly, we developed a complete scheme for optical CDMA, in which the keys are generated via down-conversion and the data is decoded by the inverse process of up-conversion (SFG).

# Optical Code-Division Multiple Access Using Broad-Band Parametrically Generated Light

Avi Pe'er, *Student Member, OSA*, Barak Dayan, Yaron Silberberg, *Fellow, OSA*, and Asher A. Friesem, *Life Fellow, IEEE, Fellow, OSA*

*Abstract*—A novel approach for an optical direct-sequence spread spectrum is presented. It is based on the complementary processes of broad-band parametric down-conversion and up-conversion. With parametric down-conversion, a narrow-band continuous-wave (CW) optical field is transformed into two CW broad-band white-noise fields that are complex conjugates of each other. These noise fields are exploited as the key and conjugate key in optical direct-sequence spread spectrum. The inverse process of parametric up-conversion is then used for multiplying the key by the conjugate key at the receiver in order to extract the transmitted data. A complete scheme for optical code-division multiple access (OCDMA) based on this approach is presented. The salient feature of the approach presented in this paper is that an ideal white-noise key is automatically generated, leading to high-capacity versatile code-division multiple-access configurations.

*Index Terms*—Code-division multiple access (CDMA), optical spread spectrum, parametric down-conversion, parametric up-conversion.

## I. INTRODUCTION

CODE-DIVISION multiple access (CDMA) is a well-known scheme for multiplexing communication channels that is based on the method of direct-sequence spread spectrum [1]. In CDMA, every channel is identified by a unique pseudonoise key, whose bandwidth is much larger than that of the input data. Ideally, the key should mimic the correlation properties of white noise and should be as long as possible in order to minimize the interference noise introduced by other channels; thus, a great deal of effort is invested in finding practical keys with good autocorrelation and cross-correlation properties [1].

The essential difference between CDMA and other multiplexing methods, such as time-division multiple access (TDMA) and frequency-division multiple access (FDMA), is that in CDMA, the resource allocated per channel is power, as opposed to time or bandwidth. This difference leads to three major advantages of CDMA compared with traditional multiplexing methods. First, CDMA is inherently flexible to dynamic changes in the bit rate and the quality of service [signal-to-noise ratio (SNR)] of any channel without affecting the total amount of data transmitted by all channels. If a channel is allowed to transmit more power, it can either improve the SNR or increase the bit rate of that channel. Consequently, this shared resource (power) can be dynamically allocated between the channels, and any channel can dynamically trade bit rate for signal to noise, and vice versa, at a given power. Second, CDMA is well adapted to dynamic changes of the number of simultaneously operating channels. Specifically, when one channel becomes inactive, the other channels benefit from the fact that the noise level is reduced. Thus, an allocated channel in CDMA that is not transmitting at a given time, *automatically* "frees its space" to other channels that need the bandwidth at that time, whereas in conventional methods, a costly system for dynamic allocation and compression is required in order to exploit inactive channels. Third, in CDMA all channels are equivalent, whereby the quality of service is that of the average channel, while in conventional methods, the quality of service is dictated by the worst channel. These advantages are eventually translated to an improved usage of the spectrum resource and a higher capacity for CDMA in many configurations.

It is obvious that the CDMA approach would be most attractive if it could be implemented optically. Indeed, several attempts have been made to incorporate optical CDMA (OCDMA) into optical communication networks [2]–[14]. The major obstacle is how to generate the direct-sequence key at the transmitter in order to encode the data and how to multiply the output of the receiver by the conjugate key in order to retrieve the data. Since, in current optical networks, the single-channel data rate is already close to the limit that electronic modulators and detectors can support and since the bandwidth of the key must be significantly broader than that of the data, it is impractical to generate the key and multiply by the conjugate key electronically. Thus, it is imperative to find a way to perform these operations *optically*.

This paper explains how to exploit the optical nonlinear process of parametric down-conversion in order to generate both an *ideal* key and its conjugate key, specifically, to develop a source that generates simultaneously both a broad-band white noise and its complex conjugate. Note that it is not necessary for CDMA that the key be previously known. Indeed, as long as both the key and the conjugate key are generated together, one can transmit the conjugate key along with the data to the receiver (at the cost of half the bandwidth). Then, the process of parametric up-conversion (also known as sum-frequency generation) is exploited to optically multiply the key and its conjugate key at the receiver. The processes of parametric down-conversion and up-conversion occur in any medium

Manuscript received November 10, 2003; revised February 12, 2004. This work was supported in part by the Yeda Research and Development Company, Limited, The Weizmann Institute of Science, Rehovot, Israel.

The authors are with the Department of Physics of Complex Systems, Weizmann Institute of Science, 76100 Rehovot, Israel (e-mail: avi.peer@weizmann.ac.il; barak.dayan@weizmann.ac.il; yaron.silberberg@weizmann.ac.il; asher.friesem@weizmann.ac.il).

Digital Object Identifier 10.1109/JLT.2004.827661

with second-order $(\chi^{(2)})$ nonlinearity. A common family of such media is that of nonlinear crystals, for example, $BaB_2O_4$ (BBO), $LiNbO_3$ (LN), $KTiOPO_4$ (KTP).

## II. PARAMETRIC CONVERSION PROCESSES

Parametric conversion processes were widely investigated over the last 35 years in both the classical and quantum mechanical frameworks [18]–[28]. Indeed, many devices based on these processes were developed, such as optical parametric oscillators (OPOs) and optical parametric amplifiers (OPAs).

The parametric down-conversion and up-conversion processes are depicted schematically in Fig. 1. Fig. 1(a) shows the parametric down-conversion, where energy is transferred from a high-frequency field (the pump field with frequency $\omega_p$ and wave vector $\vec{k}_p$) via the mediation of a nonlinear crystal to two lower frequency fields (the signal and idler fields at frequencies $\omega_s$ and $\omega_i$ and wave vectors $\vec{k}_s$ and $\vec{k}_i$, respectively). When the nonlinear medium is thick, this conversion occurs only if the phase-matching requirements are met. Specifically, $\omega_p = \omega_s + \omega_i$ and $\vec{k}_p = \vec{k}_s + \vec{k}_i$, i.e., both energy and momentum are conserved. In this down-conversion process, the phase of the generated signal field with respect to the pump is undefined *a priori* but will be opposite to that of the corresponding idler frequency ($\phi_i = \phi_p - \phi_s$), i.e., the signal and idler amplitudes are complex conjugates (see proof in Appendix I). Note that for a given pump frequency, there may be a broad band of signal–idler frequency pairs that fulfill the phase-matching requirement, depending on the specific dispersion characteristics of the nonlinear medium and on its thickness. The phase-matching bandwidth can reach hundreds of nanometers in the near infrared for thick crystals of up to several centimeters (as elaborated in Appendix II).

The process of parametric up-conversion is schematically depicted in Fig. 1(b). It is symmetrically inverse to the process of parametric down-conversion. Specifically, the energy is transferred from two low-frequency fields to a field at a high frequency that is equal to the sum of the two low frequencies. If this process is phase-matched, the phase of the generated field at the sum frequency is equal to the sum of the phases of the two low-frequency fields. Mathematically, this is equivalent to the statement that the complex field amplitude at the sum frequency is proportional to the *multiplication* of the complex amplitudes at the two low-frequency fields.

We note that although $\chi^{(2)}$ effects seem to be most practical for our purpose, other physical processes can also serve as the underlying mechanisms for the optical direct sequence communication scheme proposed here. In Appendix III, we generalize the discussion also to other processes, such as higher order nonlinear processes.

## III. THE OCDMA SCHEME

### A. Generation of the Direct-Sequence Key

The core of our approach for optical generation of the key lies in the special phase and amplitude relations between the optical fields that participate in the process of parametric down-conversion. Specifically, if a parametric source is broadly phase-matched and oscillates over a large bandwidth, there is no phase relation between the different signal (or idler) frequencies, so the signal (or idler) field is just a broad-band continuous-wave (CW) white noise, but since the phase of every signal frequency is opposite to that of the twin idler frequency, the signal and idler are *complex conjugates*. That is exactly what is required for the generation of an *ideal* key. To date, all other methods use complicated algorithms to predesign practical keys that approximate the characteristics of the desired ideal white-noise key [15]–[17]. These approximations are usually constrained by other design considerations, such as, the tradeoff between key length and design simplicity, yielding a nonoptimal result. In our approach, an ideal key is automatically generated via the physical process.

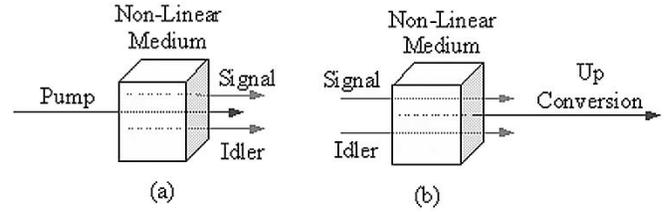

Fig. 1. Parametric processes of second-order nonlinearity. (a) Illustration of parametric down-conversion. (b) Illustration of parametric up-conversion (sum-frequency generation).

Now, the key can be externally modulated (multiplied) by the information, and both the modulated key (signal) and the conjugate key (idler) can be sent to the receiver end. Alternatively, one can modulate the narrow-band pump instead of the broad-band signal.

### B. Multiplication by the Conjugate Key at the Receiver

In order to perform the multiplication by the conjugate key at the receiver, we exploit the process of parametric up-conversion. As long as the up-conversion efficiency is low, the intensity of up-converted light at frequency $\omega$ (denoted $R(\omega)$) is given by

$$R(\omega) \propto \left| \int d\omega' A(\omega') A(\omega - \omega') \right|^2 \quad (1)$$

where $A(\omega)$ is the slow-varying amplitude of the field at frequency $\omega$. In (1), all the pairs of spectral amplitudes at frequencies that sum up to the frequency $\omega$ are added coherently. Obviously, a spectrally incoherent broad-band source yields very poor conversion efficiency, since the phases of spectral components are random, so the different pairs interfere almost destructively.

We now want to consider the up-conversion of light that was originally generated by parametric down-conversion. Since the signal and idler are complex conjugates, we can write

$$A(\omega) = A_s(\omega) + A_i(\omega) = A_s(\omega) + A_s^*(\omega_p - \omega) \quad (2)$$

where $A_s(\omega)$, the spectral amplitude of the signal field, has a random spectral phase. Assuming the spectral phases of the entire spectrum have been modulated by some general phase function $\phi(\omega)$, we obtain

$$A(\omega) = e^{i\phi(\omega)} [A_s(\omega) + A_s^*(\omega_p - \omega)]. \quad (3)$$

Inserting (3) into (1) yields

$$R(\omega) \propto \left| \int_0^{\frac{\omega_p}{2}} d\omega' \begin{bmatrix} A_s(\omega')A_s^*(\omega_p - \omega + \omega') + \\ A_s^*(\omega_p - \omega')A_s(\omega - \omega') + \\ A_s(\omega')A_s(\omega - \omega') + \\ A_s^*(\omega_p - \omega')A_s^*(\omega_p - \omega + \omega') \end{bmatrix} \times e^{i\phi(\omega') + i\phi(\omega - \omega')} \right|^2. \quad (4)$$

Equation (4) contains four terms in the integrand. Since the phase of $A_s(\omega)$ is assumed to be random, integration of the last two terms will result in a negligible contribution to $R(\omega)$ because of destructive interferences. Yet, the first two terms are basically the spectral autocorrelation of the signal field, so they have a sharp peak at $\omega = \omega_p$. The peak value is

$$R(\omega_p) \propto \left| \int_0^{\frac{\omega_p}{2}} d\omega' |A_s(\omega')|^2 e^{i\phi(\omega') + i\phi(\omega_p - \omega')} \right|^2. \quad (5)$$

The result of (5) is equal to that obtained when up-converting an ultrashort transform-limited pulse, where the phase of all frequencies is known to be zero, after applying coherent pulse shaping to it [29], [30]. Thus, we establish equivalence between the down-converted light and coherent ultrashort pulses, although the down-converted light is neither coherent nor pulsed [31].

It is clear that the up-conversion peak intensity at the original pump frequency is dramatically enhanced compared with that obtained with incoherent white noise, since all the frequency pairs that sum up to the pump frequency add *coherently*. Since the coherent integral scales linearly with the ratio of the signal bandwidth to the pump bandwidth and the incoherent integral scales only as the square root of that ratio (random walk), the enhancement $G$ in the up-conversion intensity is given by

$$G = \frac{R(\omega_p)}{R(\omega \neq \omega_p)} = \left| \sqrt{\frac{\Delta}{2\delta}} \right|^2 = \frac{\Delta}{2\delta} \quad (6)$$

where $\Delta$ is the total bandwidth of the down-converted light (signal and idler) and $\delta$ is the pump bandwidth.

In order to be able to fully utilize the enhancement in the up-conversion intensity, it is necessary to detect the up-conversion only within the pump bandwidth, filtering out the broad incoherent up-conversion spectrum. Here, the fact that the interaction occurs in a thick nonlinear medium comes to our aid. The phase-matching conditions for up-conversion in a thick medium are very strict and serve as a narrow spectral filter ($\sim$0.1-nm bandwidth at 532 nm for a 1-cm-long KTP crystal). If the detected data bandwidth is much lower, further filtering may be required. In addition, the enhancement depends critically on the characteristics of the overall phase function $\phi(\omega)$. It can be eliminated by symmetric phase functions (around $\omega_p/2$), while unaffected by antisymmetric phase functions. Thus, the application of a spectral phase function can be used to fully and reversibly control the up-conversion enhancement. For example, such a phase function can serve as a unique signature of the data-transmitting side. Only when the receiver introduces the opposite phase function is the up-conversion enhancement restored. The phase function can be as simple as a known relative delay or dispersion between the idler and the signal or an arbitrary spectral phase function introduced by the powerful tools of coherent pulse shaping [29]–[31].

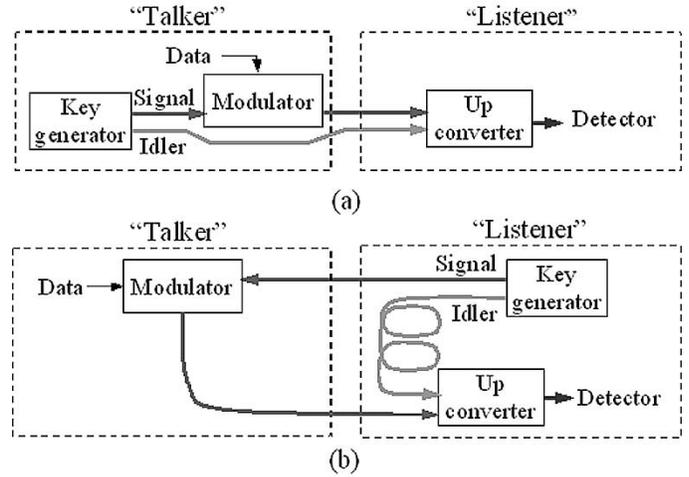

Fig. 2. Schematic configurations for optical direct-sequence communication. The key is generated at the (a) talker side and (b) listener side.

As an example, for a simple spectral phase function, consider the case of a relative delay $\tau$ between the signal and the idler fields, which is equivalent to a linear spectral phase function on half of the spectrum. In this case

$$R(\omega_p) \propto \left| \int_0^{\frac{\omega_p}{2}} d\omega' |A_s(\omega')|^2 e^{i\omega'\tau} \right|^2. \quad (7)$$

As evident from the equivalence to an ultrashort pulse, when the relative delay exceeds the coherence time of the signal field ($\tau \sim 2/\Delta$), up-conversion is dramatically reduced.

We conclude that detection of the up-conversion intensity at the original pump frequency serves as the mechanism for multiplication by the conjugate key at the receiver. The enhancement in the up-conversion intensity occurs only when the signal field combines with its twin idler field, thereby extracting the transmitted data out of the noise. This enhancement can be fully and reversibly controlled by the introduction of a spectral phase function.

### C. Single-Channel Communication

In our approach, once broad-band down-conversion is achieved, the signal field is separated from the idler field by means of a spectral filter. We identify the signal field as the key and the idler field as the conjugate key. The main difference between our approach to others in optical and electrical communications is that in our approach, the key is inherently unknown. Thus, the conjugate key must somehow reach the listener (the data-receiving side) together with the data-carrying key in order to enable the extraction of the transmitted data.

Fig. 2 shows schematically two possible communication configurations. The first probably most intuitive configuration, shown in Fig. 2(a), is that the talker (the transmitting side) generates the key and the conjugate key, modulates one of them with data, and sends both of them to the listener. The listener will perform up-conversion and extract the data. Another configuration, shown in Fig. 2(b), is a bit similar to public key encryption. The listener generates the key and the conjugate key and sends only one of them to the talker. The talker modulates the key (probably together with other noises that arrive with the data) and sends it back to the listener. The listener then uses the other key that she previously kept (properly delayed) in order to extract the data via up-conversion.

Note that any type of modulation (amplitude, frequency, or phase) is suitable for data transmission. This statement is obvious in the case of amplitude modulation (AM) but is a bit surprising for frequency modulation (FM) or phase modulation (PM), since minute frequency/phase shifts cannot be detected directly from the broad-band incoherent key (signal). Still, since the up-conversion appears at the sum frequency with a phase that is a sum of the signal and idler phases, a small frequency/phase shift of the signal will cause the same frequency/phase shift of the narrow-band coherent up-converted field, which is easily detected.

### D. Optical Code-Division Multiplexing

Our approach for code-multiplexing channels stems from the fact that enhancement in the up-conversion intensity can be controllably and reversibly destroyed by introduction of a spectral phase function; for example, by the introduction of a known amount of relative delay or material dispersion between the signal and the idler fields. This leads us to the multiplexing/demultiplexing configurations depicted in Fig. 3. In the multiplexer configuration shown in Fig. 3(a), every channel modulates the signal field emitted from a broad-band down-conversion source and introduces a unique amount of delay between the signal and the idler. The difference between the unique delays associated with the various channels should be longer than the correlation time of the broad-band signal–idler fields. All channels are then joined together in one fiber and transmitted to the receiver end.

In order for a specific receiver to detect a specific channel, the receiver configuration, shown in Fig. 3(b), will insert the inverse delay to that channel. This will restore the phase relations of this channel only, thus restoring the enhancement in the up-conversion at the original pump frequency. When the up-conversion intensity at the pump frequency is detected, this channel will be prominent above the noise (generated by other channels). Since each receiver detects only one channel and other channels just pass through without disturbance, it is reasonable that after detection the receiver will reinsert the delay in order to leave the situation unchanged for all other channels. A complete demultiplexer will be composed accordingly of many such receivers in cascade.

In our OCDMA scheme, the number of simultaneous channels, each of bandwidth $\delta$ that can be accommodated within a total bandwidth $\Delta$ is given by

$$N = \frac{1}{2}\frac{1}{\frac{s}{n}}\left(\frac{\Delta}{\delta}\right) \quad (8)$$

where $s/n$ is the minimum allowed SNR, and the major noise source is assumed to be interference caused by other channels.

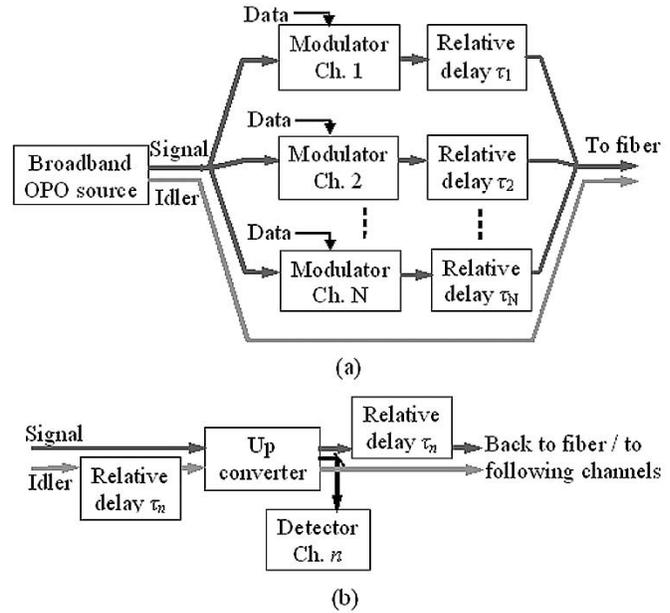

Fig. 3. Schematic configuration of an OCDMA multiplexer and demultiplexer in the case of key generation at the "talker" side. (a) Configuration of an OCDMA multiplexer. (b) Configuration of a receiver for one OCDMA channel.

This result is just a factor of two less than expected in an ideal asynchronous CDMA system, which is due to the fact that in our configuration the key has to be transmitted also. This indicates that the spectral efficiency in our approach can reach $0.5/(s/n)$. It is important to note that since the DS key in our approach is ideal (true white noise), this result is independent of practical constraints, such as lower/upper bounds on the single-channel bit rate.

Note, that in order to achieve the capacity of (8), it is necessary that all channels will share the same key, i.e., either all channels use the same signal and idler emitted by one broad-band down-conversion source or all the sources of all channels are seeded by one noise field. Although a configuration in which every channel has its own independent key is also possible, it suffers from a higher noise level at the receiver due to the existence of multiple keys in addition to multiple channels and, therefore, can support a lower number of channels $(\sqrt{N})$ compared with the number of channels $(N)$ supported by the first optimal configuration.

The multiplexing configuration of Fig. 3 is a generalization of the configuration shown in Fig. 2(a) with the keys generated by the "talker." A similar generalization can be made for the case of key generation by the "listener," as schematically shown in Fig. 4. In this case, the network contains a forward channel for all the "public keys" and a backward channel for returning data. As shown, a user who wishes to receive data (listener) generates his own keys and sends his "public key" into the forward channel of the network. Other users who wish to communicate with that user (talkers) split a part from the forward channel, modulate it with data, and insert it into the backward channel. A talker can also add a spectral phase signature to identify this specific talker–listener connection, but this is not necessary since the distance to that specific talker already serves as a unique delay sig-nature. Listeners will extract the data via up-conversion using their conjugate "private key" after appropriate delay and insertion of the opposite spectral phase. The total capacity remains as given by (8).

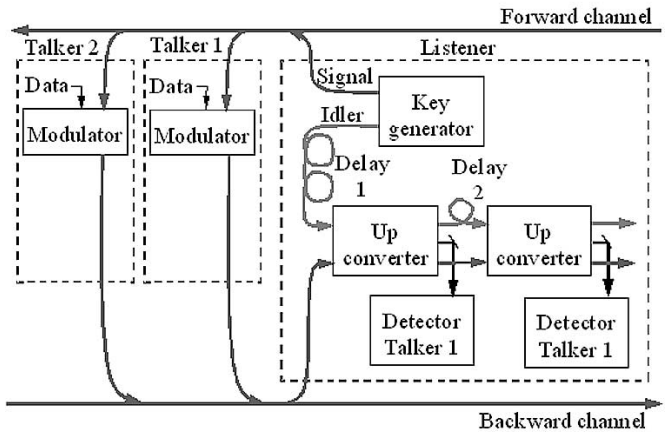

Fig. 4. Schematic configuration of an OCDMA network in the case of key generation at the "listener" side.

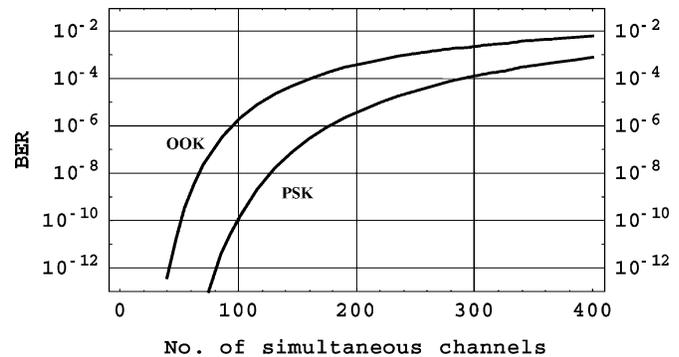

Fig. 5. Calculated BER as a function of the number of simultaneous CDMA channels for OOK and PSK modulations. The single-channel rate is taken to be 1 Gb/s with 4 THz of usable bandwidth for CDMA. The interchannel interference is assumed to be the major noise source and is modeled as a white noise, which is reasonable for a large number of simultaneous channels. No forward-error correction is included.

## IV. DISCUSSION

### A. Expected Bit-Error Rate and Data Modulation Possibilities

Equation (8) reflects the relation of the noise level at the receiver of a single channel to the total number of simultaneous channels. In Fig. 5, this relation is translated into bit-error rate (BER) for ON–OFF keying (OOK) modulation with incoherent detection and for phase-shift keying (PSK) with coherent detection. As evident from Fig. 5, coherent PSK yields much better BER results compared with incoherent OOK. This is no surprise, since coherent modulation is known to be preferable in many aspects (BER, power management, reduced nonlinearities, etc.). Albeit, coherent modulation is considered impractical in the optical domain because it requires a local oscillator at the receiver that is phase-locked to the transmitting laser. With our CDMA approach, a reference local oscillator can be easily sent from the transmitter to the receiver. Since all channels share the same down-converted field and the same pump, the pump serves as a local oscillator for all the channels. The pump cannot be sent along with the data as is (mostly because of nonlinearities), but if we "sacrifice" one channel and not modulate it with data, the up-conversion field of that channel will be just a replica of the CW pump field, which can be used as a phase reference for all other channels. Thus, coherent detection can be performed at the price of one less data channel.

Coherent modulation is made possible here due to the fact that all the channels are *optically equal* in the sense that all share the same pump (or up-conversion) and down-converted fields. This fact can be exploited for performing other kinds of optical manipulations. For example, one can think of an optical switching scheme where, after demultiplexing (up-conversion with the appropriate delay for each channel), it is possible to remultiplex the channels in a different order by down-converting them again and rearranging their delays.

### B. Privacy Considerations

Generally, spread spectrum communication is considered to be immune to both eavesdropping and jamming. Immunity to eavesdropping comes in two levels: one is the ability to conceal the mere existence of the transmittance in the noise, and the other is the encryption supplied by the use of the key.

The ability to conceal transmission in the noise is preserved in our optical direct-sequence approach, but it is clear that the encryption is lost when the talker generates both keys and sends both to the listener. Indeed, any eavesdropper can listen to the transmission if its existence is known to him or her just by performing up-conversion. The encryption can be restored if the talker introduces an additional overall spectral phase function known only to him or her and the listener, so only the listener can undo this phase function before performing up-conversion.

The "public key" communication possibility seems at first glance to be totally immune to eavesdropping. The conjugate key that is never transmitted is inherently unknown and is never repeated (it is continuously generated) so that an eavesdropper cannot reproduce the conjugate key. Although this statement is correct, an eavesdropper can still listen in on the data transmitted without having the other key. This is due to the fact that the talker does not know the key either and modulates data onto whatever is received within the spectral bandwidth of the key. Thus, if an eavesdropper knows about the communication, she can generate her own set of keys and send one of these keys also to the talker, who will modulate this key also with data. The eavesdropper can now extract the data by up-conversion using her own properly delayed conjugate key. Again, talkers can prevent eavesdropping by introducing a unique spectral phase function known only to them and the listener.

We conclude that the fact that the keys are incoherent and inherently unknown cannot provide additional security to the communication with the previously described schemes, so in order to achieve security of the transmission, one must "encrypt" the key by use of a unique spectral phase function.

### C. Chromatic Dispersion

Dispersion is an important optical factor that influences data transmission over a fiber. Generally, the proposed optical DS

is expected to be sensitive to dispersion because of the large bandwidths of the signal and the idler and the need to keep their phases intact. This is true for *even* orders of dispersion, and the total usable bandwidth for CDMA communication is limited by the degree of compensation for the even orders; therefore, a good compensation for the second-order dispersion ($\beta_2$) is necessary. Yet, as long as the pump bandwidth (i.e., the data bit rate per channel) is not very high, the proposed OCDMA is insensitive to odd orders of dispersion. This is due to the fact that odd orders introduce a spectral phase, which is antisymmetric around the center of the spectrum ($\omega_p/2$), and as clear from (5), the up-conversion intensity is insensitive to it.

As the pump bandwidth (or the single-channel data rate) gets higher, the center frequency is less defined, and odd orders of dispersion start to manifest themselves as residual even orders at the side lobes of the pump. Thus, the even orders of dispersion limit the total usable bandwidth for CDMA, while the odd orders of dispersion limit the single-channel bandwidth.

A numerical calculation using the dispersion parameters of bulk silica shows that after propagating 40 km at the zero-dispersion point, where the second order of dispersion is zero, the total usable bandwidth is ∼5 THz (i.e., ∼2.5 THz available for data transmission), and the maximal bit rate per channel is ∼10 GHz. Note that the limit posed by odd orders of dispersion on the single-channel bit rate does not affect the total capacity of the CDMA connection since, with CDMA, a reduction of the single-channel bandwidth automatically increases the total number of channels, leaving the capacity unchanged.

### D. Nonlinear Propagation Effects

Nonlinear effects, such as Brillouin scattering, self-phase modulation, cross-phase modulation, and four-wave mixing, occur when the temporal intensity or the power spectrum in the fiber is high; thus, they are generally minimized by broad-band noiselike fields that minimize the possibility for persisting constructive interferences. It is obvious that once the signal and the idler are "decorrelated" by introduction of some spectral phase function (e.g., a short relative delay), they can be considered as a broad-band noise field, so our OCDMA scheme inherently minimizes nonlinear effects in fibers that can have unwanted influence on the communication.

### E. Comparison With Other OCDMA Approaches

The major problem for obtaining OCDMA is that of generating the pseudonoise key. The many attempts to solve this problem can be divided into two categories—a coherent approach and an incoherent approach. The coherent approach [2]–[6] starts from a broad-band coherent source (i.e., a mode-locked laser that emits transform-limited pulses). The key for each channel is then generated by actively shaping the spectral phase in a unique manner through some kind of a pulse-shaping device, which deforms the pulse to mimic a pseudonoise burst. At the receiver, a shaper performs the inverse shaping to recreate the original transform-limited pulse, which is then detected by use of an ultrafast nonlinear thresholder that is sensitive to the pulse peak intensity. The main disadvantage of this approach is that a great deal of the flexibility of CDMA is lost because of the limitations imposed by active pulse shaping. For example, the total number of channels is limited by the number of pixels of the pulse shaper, and the lowest effective bit rate per channel is limited by the spectral resolution of the shaper.

The incoherent approach (with its many versions) [7]–[14], on the other hand, involves an incoherent broad-band source. Although such a source emits "true noise," the phase of the emitted field is not known, so only intensity manipulations are possible. This makes the incoherent approach robust in the sense that it is relatively immune to phase changes due to propagation effects, but since the incoherent approach is inherently unipolar, the cross correlation of different keys cannot average out to zero. Thus, the noise level in incoherent CDMA systems is much higher, which causes the performance to deteriorate severely as the number of users grows [3], [11]. For this reason, the capacity of incoherent CDMA systems is inherently and significantly lower than that of coherent ones ($\sqrt{N}$ channels compared with $N$ channels in the coherent approach).

Our OCDMA approach can be considered as some kind of a hybrid that alleviates some limitations of both approaches. It is a coherent approach in the sense that it relies on the coherent phase relation between the signal and the idler (the key and its conjugate) so that the capacity is comparable with that of the coherent approach. Yet, the key is a true white noise that is passively generated, minimizing nonlinear effects and preserving the full flexibility of CDMA. The only drawback of the proposed OCDMA compared with the coherent approach is the need to transmit the key to the receiver, whereby 50% of the bandwidth is "wasted." Fortunately, even this drawback has a positive side because it leads to the insensitivity to odd orders of dispersion.

### V. Concluding Remarks

In this paper, we presented novel coherent optical direct-sequence communication and CDMA configurations that exploit the special spectral phase correlations within the spectrum of broad-band parametrically down-converted light in order to generate an *ideal* set of key and conjugate key. Extraction of the encoded data is performed by the inverse process of parametric up-conversion. The proposed scheme benefits from the high capacity associated with coherent optical processing, while avoiding the limitations of active pulse shaping, thereby preserving the full flexibility of CDMA.

For the proposed optical CDMA schemes to be practical, it is obviously necessary to have a parametric source that generates the broad-band signal and idler fields with high efficiency at low pump powers. Since the nonlinear interaction is weak, an oscillator in a high-finesse cavity would appear to be a reasonable choice. Unfortunately, in such a cavity, mode competition significantly narrows the spectrum of the actual oscillation even when phase matching allows broad oscillation. Consequently, in order to achieve broad-band oscillation, the cavity should be specially designed to suppress narrow-band oscillation while allowing broad-band oscillation to develop. We are currently investigating such a source and expect to report on the design, performance analysis, and experimental results in the near future.

## APPENDIX I
## PHASE AND AMPLITUDE CORRELATIONS IN PARAMETRIC DOWN-CONVERSION

In order to understand the signal–idler correlation in parametric down-conversion, we adopt the theoretical treatment developed for three-wave mixing that appears in [18, Ch. 6, pp. 67–85]. We start from the standard equations of three-wave mixing describing the down-conversion process. Under the simplifying assumptions of a lossless medium and perfect phase matching, we have

$$\frac{\partial A_s}{\partial z} = -i\kappa A_i^* A_p$$
$$\frac{\partial A_i}{\partial z} = -i\kappa A_s^* A_p$$
$$\frac{\partial A_p}{\partial z} = -i\kappa A_s A_i \quad (9)$$

where $A_s$, $A_i$, and $A_p$ are the slow-varying amplitudes of the signal, idler, and pump, respectively, and $\kappa$ is the nonlinear coupling, which is related to the nonlinear coefficient $d$ (in MKS units) via

$$\kappa = \frac{d}{2}\sqrt{\frac{\mu_0}{\varepsilon_0}\frac{\omega_s\omega_i\omega_p}{n_s n_i n_p}} \quad (10)$$

where $n_x$ is the refractive index and $\omega_x$ the frequency of field $x$ ($x = s, i, p$). $\varepsilon_0$ and $\mu_0$ are the vacuum dielectric permeability and magnetic permeability, accordingly. Our interest is in the phase correlations between the three amplitudes, so a transformation to polar coordinates seems reasonable

$$\frac{\partial R_s}{\partial z} + iR_s\frac{\partial \theta_s}{\partial z} = -i\kappa R_i R_p \exp\left[i(\theta_p - \theta_s - \theta_i)\right]$$
$$\frac{\partial R_i}{\partial z} + iR_i\frac{\partial \theta_i}{\partial z} = -i\kappa R_s R_p \exp\left[i(\theta_p - \theta_s - \theta_i)\right]$$
$$\frac{\partial R_p}{\partial z} + iR_p\frac{\partial \theta_p}{\partial z} = -i\kappa R_i R_s \exp\left[-i(\theta_p - \theta_s - \theta_i)\right] \quad (11)$$

where we substituted $A_x = R_x \exp(i\theta_x)$ for all three waves. We now substitute $\Delta\theta = \theta_p - \theta_s - \theta_i$ into (11) and separate the real and imaginary parts to yield

$$\frac{\partial R_s}{\partial z} = \kappa R_i R_p \sin\Delta\theta$$
$$\frac{\partial R_i}{\partial z} = \kappa R_s R_p \sin\Delta\theta$$
$$\frac{\partial R_p}{\partial z} = -\kappa R_s R_i \sin\Delta\theta$$
$$\frac{\partial \Delta\theta}{\partial z} = \kappa\cos\Delta\theta\left[\frac{R_i R_p}{R_s} + \frac{R_s R_p}{R_i} - \frac{R_i R_s}{R_p}\right]. \quad (12)$$

Substituting the first three parts of (12) into the fourth and performing some simple algebraic manipulations yields

$$\frac{\sin\Delta\theta}{\cos\Delta\theta}\frac{\partial \Delta\theta}{\partial z} = \frac{1}{R_s}\frac{\partial R_s}{\partial z} + \frac{1}{R_i}\frac{\partial R_i}{\partial z} + \frac{1}{R_p}\frac{\partial R_p}{\partial z}. \quad (13)$$

Equation (13) is equivalent to

$$-\frac{\partial}{\partial z}\left[\ln(\cos\Delta\theta)\right] = \frac{\partial}{\partial z}\left[\ln(R_s R_i R_p)\right] \quad (14)$$

which can be solved immediately to obtain

$$\cos\Delta\theta = \frac{C_1}{R_s R_i R_p} \quad (15)$$

where $C_1$ is an integration constant. Since the phase difference $\Delta\theta$ is real, it is clear that the constant $C_1$ is bound by the initial values of the field amplitudes $R_x[0]$, according to

$$0 \leq |C_1| \leq R_s[0]R_i[0]R_p[0]. \quad (16)$$

In most practical cases, at least one of the fields $A_s$, $A_i$, and $A_p$ is initiated by spontaneous emission noise, so it is practically zero. Thus, as the field amplitudes grow, the denominator of (15) becomes much larger than the nominator, so the value of the constant $C_1$ becomes irrelevant, and, for all practical purposes, we obtain $\cos\Delta\theta = 0$. We thus find that the phases of the signal and the idler are correlated according to

$$\theta_s + \theta_i = \theta_p - \frac{\pi}{2}. \quad (17)$$

If we choose to define the pump phase as $\theta_p = \pi/2$, we get

$$\theta_s = -\theta_i. \quad (18)$$

Consequently, we find that the phase of an idler mode in an OPO cavity is opposite to that of the corresponding signal mode. The absolute value correlation between the signal and the idler can be understood from the fact that they are symmeftric in (11). Thus, if the initial conditions are symmetric, then this symmetry will be preserved.

## APPENDIX II
## BROAD-BAND PHASE MATCHING

In Section III-B, we showed that the temporal resolution obtained by a parametric source is equal to that obtained by a transform-limited pulse. Consequently, if we intend to explore and exploit this feature, we should develop an OPO/OPA oscillating over the widest possible spectrum. A necessary condition for a broad-band oscillation is phase matching over a broad wavelength range. It is generally known that when the signal and the idler are close to degeneracy (i.e., $\omega_i \approx \omega_s \approx \omega_p/2$), the type I phase matching (where the signal and the idler have the same polarization) becomes broad. This is depicted in the graph in Fig. 6(a), which shows the phase mismatch ($\Delta k \equiv k_p - k_s - k_i$) as a function of wavelength for a periodically polled KTP crystal pumped at 532 nm. As evident, to first-order approximation in wavelength, the phase-matching condition for wavelengths close to that point is the same. One can expect a spectral width of tens of nanometers around 1064 nm for a crystal length of 1 cm.

Yet, much broader phase matching is possible if the pump is tuned so that the degeneracy point (at the wavelength of $\lambda = 2\lambda_p = 4\pi c/\omega_p$) coincides with the point of zero dispersion of the crystal. At the zero-dispersion point, the second derivative of the index of refraction with respect to the wavelength vanishes, so the index of refraction is predominantly linear in wavelength. It can be shown that when the index of refraction is linear, any two complementary wavelengths are phase matched. Higher orders of dispersion will limit the phase-matching bandwidth but only to fourth order in wavelength, since odd orders of dispersion do not affect phase matching. Thus, with zero dispersion, one can obtain ultrabroad phase matching, of up to hundreds of nanometers, as depicted in Fig. 6(b) for a BBO crystal pumped at 728 nm.

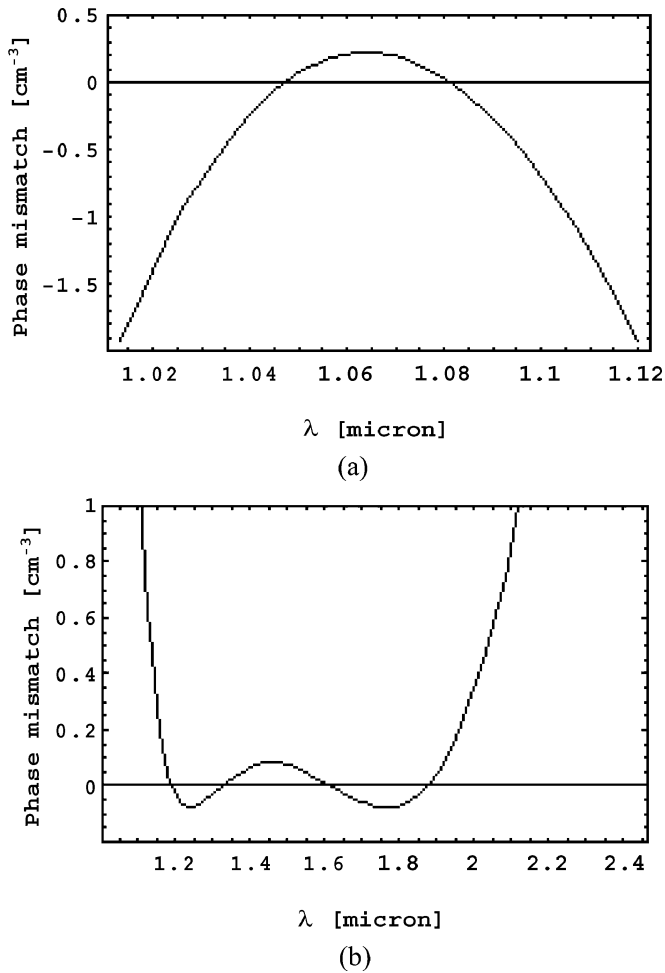

Fig. 6. Broad type I phase-matching situations. (a) Typical phase mismatch as a function of wavelength near the degeneracy point (calculated here for a periodically polled KTP crystal pumped at 532 nm). (b) Ultrabroad situation when the pump is tuned such that the down-conversion center wavelength coincides with the zero-dispersion wavelength of the crystal (as calculated here for a BBO crystal pumped at 728 nm).

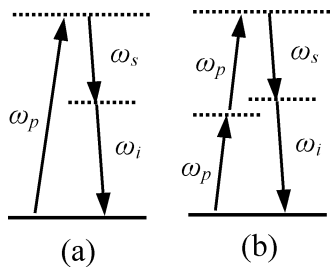

Fig. 7. Feynman diagrams of possible nonlinear mechanisms for CDMA. (a) Three-wave mixing. (b) Four-wave mixing.

## APPENDIX III
## USING HIGHER ORDER NONLINEAR PROCESSES

In the text, we considered a CDMA scheme based on the process of three-wave mixing as the mechanism for generating two broad-band conjugate fields. This process can be schematically depicted by the Feynman diagram shown in Fig. 7(a), which can be interpreted quantum mechanically as the conversion of one high-energy pump photon into two low-energy signal and idler photons. Yet, higher order nonlinear processes can perform this task just as well, e.g., with four-wave mixing, as depicted in Fig. 7(b), which can be interpreted as the conversion of two pump photons into two broad-band signal and idler photons. Generally, any process in which $n$ pump photons are converted into two photons can be considered as the basis for our CDMA scheme, though it is unlikely that such higher order nonlinearities will be preferable because of their weakness and added complexity.

**Avi Pe'er** was born in Petah Tiqwa, Israel, in 1970. He received the B.Sc. degree in physics and computer science from Tel-Aviv University, Tel Aviv, Israel, in 1996 and the M.Sc. degree in physics from the Weizmann Institute of Science, Rehovot, Israel, in 1999 with a thesis on optical processing with totally incoherent light. He is currently working toward the Ph.D. degree in physics at the Weizmann Institute of Science, with a dissertation on the field of nonlinear optics, focusing on the special characteristics of broad-band down-converted light and its applications.

Mr. Pe'er is a Student Member of the Optical Society of America (OSA).

**Barak Dayan** was born in Tel Aviv, Israel, in 1970. He received the B.Sc. degree in physics and mathematics and the M.Sc. degree in physics, both from the Hebrew University, Jerusalem, Israel, in 1991 and 1998, respectively. The M.Sc. thesis involved holography and spectral analysis of chromosomes. He is currently working toward the Ph.D. degree in physics at the Weizmann Institute of Science, Rehovot, Israel, with a dissertation focusing on experimental quantum optics and coherent control.

**Yaron Silberberg** received the Ph.D. degree in applied physics from the Weizmann Institute of Science, Rehovot, Israel, in 1984.

He joined Bell Communications Research (Bellcore), Holmdel, NJ, in 1984, first as a Postdoctoral Researcher and then as a regular Member of the technical staff. At Bellcore, he studied, among other things, linear and nonlinear guided-wave optics, optical amplifiers, and soliton physics. In 1994, he joined the Weizmann Institute of Science as an Associate Professor, where he established a new laboratory in ultrafast optics. In 1999, he was promoted to Professor and also was appointed head of the Department of Physics of Complex Systems. Since 2002, he has been the Dean of the Faculty of Physics at the Weizmann Institute of Science. The main topics he has studied at Weizmann are nonlinear microscopy, ultrafast quantum coherent control, and nonlinear phenomena in waveguides and fibers.

Prof. Silberberg is a Fellow of the Optical Society of America (OSA) and, among other professional services, he served as the Chair of the Gordon Conference in Nonlinear Optics and Lasers in 1995.

**Asher A. Friesem** (S'57–M'62–SM'79–F'95–LF'02) received the B.Sc. and Ph.D. degrees from the University of Michigan, Ann Arbor, in 1958 and 1968, respectively.

He was with Bell Aero Systems Company and Bendix Research Laboratories from 1958 to 1963. From 1963 to 1969, he conducted investigations in coherent optics, mainly in the areas of optical data processing and holography, at the University of Michigan's Institute's of Science and Technology. From 1969 to 1973, he was Principal Research Engineer in the Electro-Optics Center of Harris, Incorporated, performing research in the areas of optical memories and displays. He joined the staff of the Weizmann Institute of Science, Rehovot, Israel, in 1973 and was appointed Professor of Optical Sciences in 1977. He subsequently served as Department Head, Chairman of the Scientific Council, and Chairman of the Professorial Council. Over the years, he has been a Visiting Professor in Germany, Switzerland, France, and the United States. He has authored and coauthored 200 scientific papers, has been coeditor of three scientific volumes, and holds more than 25 international patents. In recent years, his research activities have concentrated on new holographic concepts and applications, optical image processing, electrooptic devices, and new laser resonator configurations.

Prof. Friesem is a Fellow of the Optical Society of America (OSA) and a Member of the International Society for Optical Engineers (SPIE) and Sigma Xi. He has served on numerous program and advisory committees of national and international conferences. Among other posts, he currently serves as a Vice President of the International Commission of Optics (ICO) and is also Chairman of the Israel Laser and Electro-Optics Society.

# Chapter 4
# Quantum Lithography
# by Coherent Control of Classical Light Pulses

Quantum lithography is a method that was recently suggested to overcome the standard Rayleigh diffraction limit and improve the resolution of optical lithography by exploiting non-classical entangled states of the electromagnetic fields [41-43]. The method utilizes interference between groups of $N$ entangled photons to enable the registration of spots smaller by a factor $N$ than classicaly possible.

The research of the two-photon coherence properties of high power down converted light led us to the understanding that non-classical states of the light are not necessary for this purpose. Specifically, we investigated a simple method to obtain $N$ photons interference with classical pulses that excite a narrow multiphoton transition, thus shifting the "quantum weight" from the field to the lithographic material [44]. A novel practical lithographic scheme was developed and the underlying principles were demonstrated by a two-photon interference experiment in atomic Rubidium, where focused spots beat the diffraction limit by a factor of 2.

# Quantum lithography by coherent control of classical light pulses

Avi Pe'er, Barak Dayan, Marija Vucelja, Yaron Silberberg and Asher A. Friesem

*Department of Physics of Complex Systems, Weizmann Institute of Science, Rehovot 76100, Israel*
*avi.peer@weizmann.ac.il*

**Abstract:** The smallest spot in optical lithography and microscopy is generally limited by diffraction. Quantum lithography, which utilizes interference between groups of *N* entangled photons, was recently proposed to beat the diffraction limit by a factor *N*. Here we propose a simple method to obtain *N* photons interference with classical pulses that excite a narrow multiphoton transition, thus shifting the "quantum weight" from the electromagnetic field to the lithographic material. We show how a practical complete lithographic scheme can be developed and demonstrate the underlying principles experimentally by two-photon interference in atomic Rubidium, to obtain focal spots that beat the diffraction limit by a factor of 2.

**OCIS codes:** (110.5220) Photolithography, (020.4180) Multiphoton processes, (320.7110) Ultrafast nonlinear optics, (180.0180) Microscopy,

Over the last several decades, optical lithography has evolved into a major technique of the semiconductor industry (and elsewhere) for writing fine features on a surface. Due to the high demand for miniaturization, great effort has been invested in obtaining the highest resolution possible in practical lithographic schemes [1]. The minimal feature size that can be written / imaged in linear optics lithography is limited by diffraction to $d_1 \sim \lambda/NA$, where $\lambda$ is the optical wavelength and *NA* is the numerical aperture of the imaging setup [2]. Essentially, the minimal spot is generated by the interference between many plane waves arriving to the substrate at different angles (as high as allowed by the numerical aperture).

The use of non-linear *N*-photon absorption in the lithographic medium leads to an improvement of factor $\sqrt{N}$ in the spot size ($d_N = d_1/\sqrt{N}$) due to the enhanced contrast of the material response, as was indeed demonstrated [3]. Further improvement of *N*-fold over the diffraction limit was suggested by use of "Quantum Lithography" [4-6]. Unlike standard optical lithography, where the single photon acts as the interfering entity, in quantum lithography the interference is between groups of *N* entangled photons that arrive at different angles. This multi-photon interference is equivalent to single-photon interference with a wavelength of $\lambda/N$, leading to the desired improvement in resolution. In order to obtain a spot size $d_1/N$ with quantum lithography, two conditions must be met: First, the absorption in the lithographic material should be purely *N*-photonic. Second, the photons should interfere only as *N*-photons entangled groups, but not as single photons; i.e. the absorption of the *N* photons should be either from one direction or another, excluding possibilities of absorbing some photons from one direction and the rest from another [5, 7]. This apparently implies the need for either time-energy or momentum-space entanglement. Quantum entanglement is not easily generated or maintained. Specifically, a severe inherent power limit is imposed if the light is to be considered as a flux of entangled photons (intuitively, the flux must be low enough that the entangled *N*-photon groups arrive "one at a time") [7]. This need to operate at the single-photon regime severely restricts the practicality of quantum lithography.

In this paper we demonstrate that field quantization effects are not necessary and can be replaced by the quantum nature of the lithographic material (i.e., a spectrally narrow *N*-photons transition). As a result, higher powers can be utilized according to standard practical constraints without any inherent limitation. The main principle in our approach is that when a short pulse excites a narrow transition in the material, the excitation lifetime is much longer than the pulse duration. Thus, if the transition is excited again by another pulse within the excitation lifetime of the first pulse, these two excitations can interfere even if the two pulses do not (i.e. are mutually incoherent). Since this interference occurs through the medium, the relative phase that affects it is dictated by the transition frequency $\omega_A$ and the relative delay $\tau$ between the pulses ($\phi = \omega_A \tau$). If this excitation is non-linear of order *N*, the center frequency of the exciting pulse is $\omega_0 = \omega_A/N$. As a result, this "quantum interference" is equivalent to one-photon interference with a wavelength shorter by a factor of *N*, as was indeed demonstrated for *N*=2 [8]. This long lived interference through the medium, between very short pulses is the classical equivalent of time-energy entanglement in quantum lithography; the narrow bandwidth of the transition imposes *N*-photon coherence between mutually incoherent (non-overlapping) pulses. In the following we theoretically analyze and experimentally demonstrate how control over the spatial dependence of the "quantum phase" $\phi(r) = N\omega_0\tau(r)$ leads to a complete lithographic scheme with a resolution that is improved by a factor of *N*.

A basic question in lithography is how to generate the smallest possible spot with a given lithographic lens. We wish to answer this question for our quantum lithography scheme. The analysis given here assumes one transverse dimension for simplicity (a generalization to two dimensions is straightforward). In standard lithography a small spot is achieved simply by focusing a plane wave with a lens, generating a spot of size $d_1 \approx \lambda f/D$, where f is the focal length of the lens, D is the beam diameter and paraxial optics is assumed. We wish to obtain an additional factor of *N* in resolution using quantum interference of a train of pulses.

Mathematically, the interfering entity is the amplitude of the non-linear response in the lithographic material. In the absence of intermediate resonant transitions this amplitude is proportional to the resonant frequency component of the *N*th power of the electric field $E^N(\omega_A)$ [9, 10]. Assuming for simplicity two exciting pulses ($E_1$ and $E_2$), the intensity of excitation will be

$$I(x) \propto \left| E_1^N(x;\omega_A) + E_2^N(x;\omega_A) \right|^2, \qquad (1)$$

where *x* is the spatial coordinate. Note that since the pulses do not overlap in time, mixed terms (e.g., $E_1^p E_2^{N-p}$) are absent. While this expression assumes a weak field perturbative regime, extension to stronger fields is possible [10, 11] without change to the relevant features.

The key element in achieving a small spot is to shape the spatial phase fronts of the exciting pulses at the focus such that they will interfere constructively in the center of the spot and destructively near the edges. Specifically, we are searching for *M* pulse fields at the lens surface $\{\varepsilon_k(x_l)\}$ whose corresponding focal fields $\{E_k(x_f)\}$ fulfill

$$\left| \sum_{k=1}^{M} E_k^N(x_f) \right|^2 = I(x_f), \qquad (2)$$

where $I(x_f)$ is the desired narrow lithographic spot. Two constraints must be considered. First, the focal spots cannot be smaller than the diffraction limit imposed by the aperture of the lens. Second, due to the non-linearity, spreading the energy either in space or in time is not desired, so the number of pulses *M* and the spatial extent of the focal fields should be minimized (i.e., maximize the spatial extent at the lens). While a general solution is currently unknown to us, a simple practical approach is to divide the lens into two non overlapping segments and delay the pulse in one of the segments (e.g., with a piece of glass) as schematically sketched in Fig. 1. Since each segment can be considered as an off axis lens, the two segments generate overlapping spots with linear phase fronts of opposite slopes. If the delay between the pulses is tuned correctly, this will lead to the desired constructive interference at the center. It is interesting to note that a Fourier equivalent of the segmentation scheme of Fig. 1 was suggested for doubling the resolution with two-photon absorption [12, 13], where different *spectral* components are separated spatially to two segments on the lens. While such an approach can yield similar results for the two-photon case, generalization to the *N* photon case is not straight forward and its implementation is more complicated.

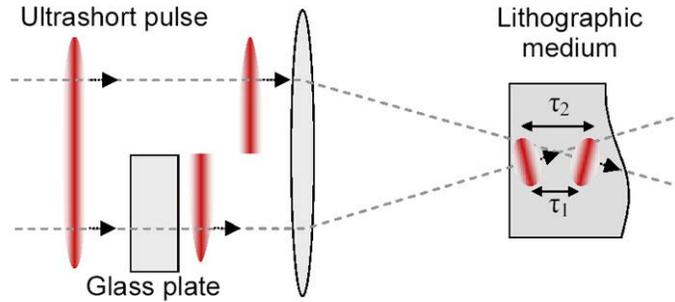

Fig. 1. Schematic setup for generation of sub diffraction limited spots by quantum interference. A glass plate delays half of a planar ultrashort pulse with respect to the other half. As a result, the non-linear lithographic medium at the focus is excited by two consecutive pulses with a space-variant relative delay; thus generating a space-dependent two-photon interference. Fine tuning the delay can be performed by a small tilt of the glass.

Using Eq. (2) we calculated the sub-diffraction limited spots for the arrangement Fig. 1, as compared to the single-photon diffraction limited spot assuming paraxial optics. The results are presented in Fig. 2. Figures 2(a) and 2(b) show the resulting focal intensity spots for two-photon absorption and four-photon absorption. As evident, the higher resolution is indeed achieved, but at the cost of side lobes, which increase for higher order non-linearity. Additional segmentation of the lens shifts the side lobes away from the main lobe [12], but does not attenuate them, as can be seen in Fig. 2(c) and 2(d). Moreover, segmentation spreads the incident pulse energy in both time and space, thus reducing drastically the total material response. A simple solution for side-lobes suppression is to add only one more pulse with appropriate phase and two regular foci (one on each side lobe), as was done to obtain the results of Fig. 2(e) and 2(f). Once a single small spot is achieved, it is clear that any pattern can be written on a substrate by scanning the spot over it. Yet, unlike linear optics lithography, our scheme is not directly suitable for imaging lithography of arbitrary patterns, since one-photon interference can cause mixing between close $N$-photon spots. However, any use of non-linearity inherently favors scanning because of power considerations. Note that even though the intensity of the non-linear lithographic response with entangled photons depends *linearly* on the incoming photon flux, the above power considerations also apply, since the sensitivity to spatial expansion remains non-linear [5, 7].

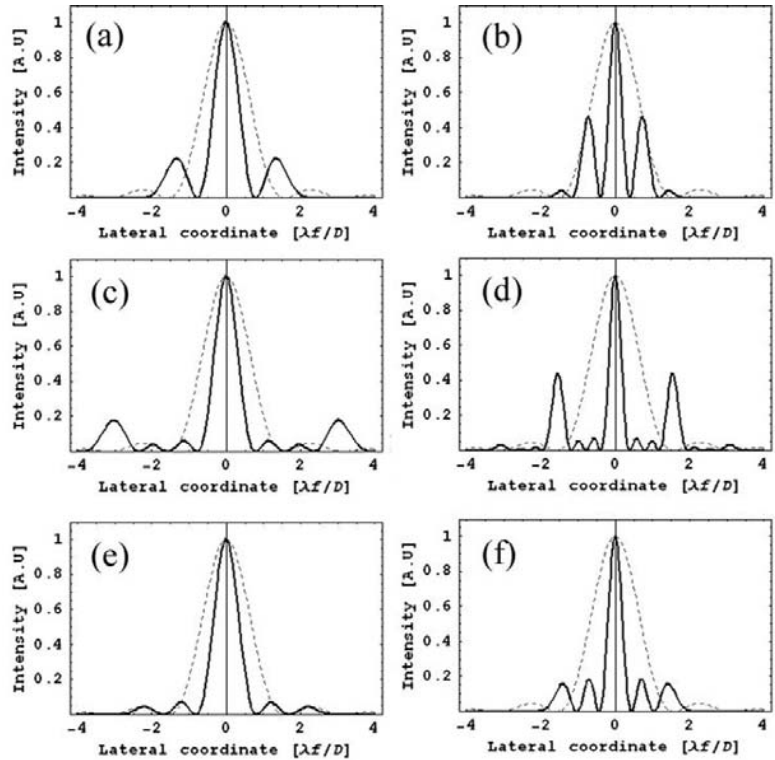

Fig. 2. Calculated sub diffraction-limit spots (line), as compared to the diffraction-limited single-photon spot (dashed). (a), (b) are the two-photon and four-photon spots respectively of the two segment configuration of Fig. 1. (c), (d) show the same spots assuming four equal non-overlapping segments instead of two. (e), (f) show these spots when a third pulse with two offset foci is used to suppress the side lobes. All segments are assumed to be illuminated by a uniform plane wave pulse.

We chose to demonstrate the principles of our quantum lithography scheme in general and the ability to generate sub diffraction-limited spots in particular, using a narrowband two-photon transition in atomic Rubidium. Our experimental configuration (shown in Fig. 3), is essentially a realization of the basic setup of Fig. 1. The cylindrical telescope weakly focuses in one dimension pulses (~100fs around 778nm) emitted from a Ti:Sapphire laser into the Rb cell. The delay line induces a variable relative delay between the two halves of the pulse. We detected the two-photon excitation by imaging the resulting fluorescence at 420nm onto an Enhanced CCD camera. Since our measurements were performed only in one dimension, the beam was tightly focused in the perpendicular dimension with a strong cylindrical lens in front of the cell in order to increase the signal. Since diffusion of the atoms in the gas during the fluorescence lifetime may blur the two-photon spot, we added a buffer gas (Neon at a pressure of 400Torr) to the Rubidium cell, thus slowing down the spread of the excited atomic cloud by elastic collisions with the buffer. In addition, we used a rather weak focusing into the cell to generate relatively large spots that were not easily washed out.

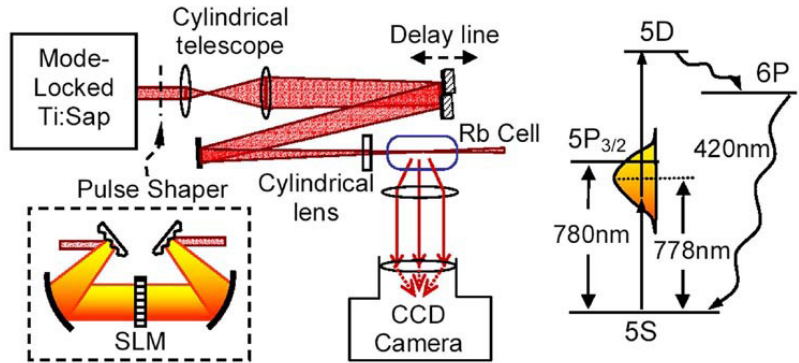

Fig. 3. Experimental configuration and relevant level diagram for atomic Rb. The cylindrical telescope weakly focuses the beam into the Rb Cell, the delay line controls the interference and the CCD records the image of the fluorescence spot. The cylindrical lens in front of the cell tightly focuses the beam in the perpendicular dimension.

We chose for our experiment the 5S-5D two-photon transition centered at 778nm (see the relevant level diagram in Fig. 3). While this transition is relatively strong, we had to avoid the excitation of the intermediate 5P level (at 780), since the resonant two-step excitation via that level has a very long lifetime that would have created undesired interference between the pulses, even at very long delays. Thus, we blocked the resonant frequency at 780nm in a pulse shaper that was placed at the entrance to the experimental configuration. We also utilized the pulse-shaper to introduce a $\pi$ phase shift to frequencies above the resonance and below it, in order to maximize the excitation [14]. Note, that in a normal situation with a non-resonant two-photon transition, the pulse shaper is not required at all.

The experimental results are shown in Fig. 4. The double resolution is demonstrated clearly in Fig. 4(a), where two CCD images and the corresponding transverse line cross sections (along with theoretical fits) are presented. The images were taken when the delay line was tuned to obtain a "dark spot" (i.e., destructive interference in the center of the spot). At a delay shorter than the coherence length of the pulse, regular one-photon interference is observed. Yet, when the delay was tuned far beyond the coherence length, two-photon interference is observed and the distance between the lobes is reduced to a half. Figure 4(b) shows a "bright spot" (constructive at the center) at the two-photon interference regime, demonstrating the two-fold narrowing of the central lobe, as well as the expected side lobes. The discrepancy between the theoretical and experimental results in Fig. 4 is mainly due to imperfect blocking of the resonant component at 780nm in our shaper that led to a residual narrowband response, and partly due to vibrations of the delay line mirrors during the CCD capture time and to diffusion of the Rb cloud during fluorescence.

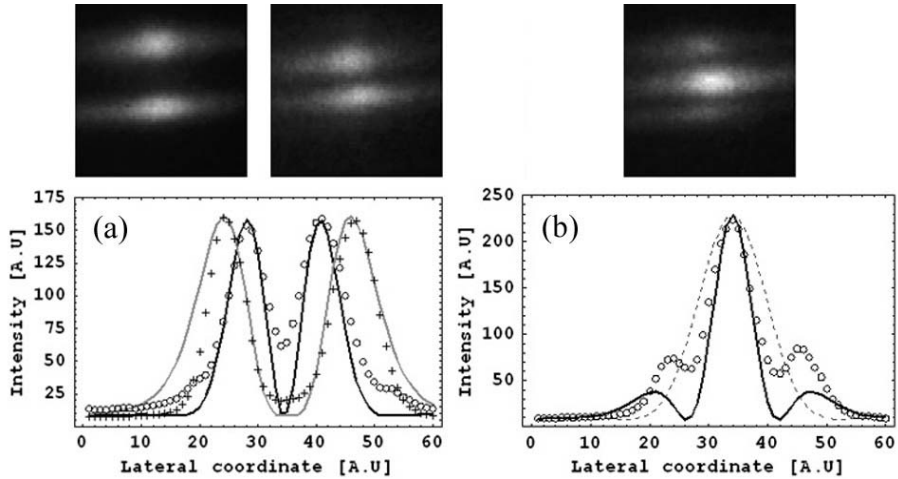

Fig. 4. Experimental results. (a) images and transverse cross sections of "dark spots" (destructive at the center) for a short relative delay (crosses – data, gray line – theoretical fit) and a long relative delay (circles – data, line - theoretical fit), demonstrating the double resolution of two-photon interference compared to one-photon interference. (b) is the corresponding two-photon "bright spot" as compared to the diffraction limited one-photon spot (dashed). All experimental cross sections were averaged over the center portion of the image (18 pixel lines) to reduce noise. The theoretical fits assume a Gaussian beam profile with a narrow gap in the middle.

It is interesting to note that while our discussion focused on coherent ultrashort pulses, a coherent pulsed excitation is not necessary. Since narrow two-photon transitions are unaffected by anti-symmetric spectral phase functions [9], broadband high-power down-converted light, which is both continuous and incoherent can be the source of illumination just as well because of it's inherent anti-symmetric spectral phase [15].

To conclude, we theoretically analyzed and experimentally verified a simple and practical scheme for sub diffraction limit quantum lithography which relies on the quantum nature of the lithographic material and not of the exciting field. In order for the method to be practical, a non-linear lithographic material with a narrow excitation line is required. We speculate that such a material should include some atomic or ionic doping that will trigger the lithographic chemical reaction. We hope that the relatively simple implementation of this method will lead to the development of such materials and to practical applications. While lithography was the major application considered here, the ability to excite sub-diffraction limit spots may be applied to non-linear fluorescence microscopy just as well.

# Chapter 5
# Pair-wise Mode-Locked OPO

In the previous chapters, the equivalence of broadband down converted light and ultrashort pulses was established and exploited for several applications. This equivalence extends also to the sources that generate the light.

In general, efficient generation of coherent radiation requires stimulation of the generating process in an optical cavity, and down conversion is no exception to this rule [45]. However, mode competition is a common problem that drastically narrows the oscillating spectrum in efficient cavities. As a result a special design is required to suppress mode competition and achieve a broadband oscillation. For ultrashort pulses a broad oscillation is achieved by the technique of passive mode locking [46-48]. We investigated the design and experimental realization of a low threshold OPO source that oscillates over the entire bandwidth allowed by phase matching. We exploited two-photon absorption as a selective loss mechanism in the cavity to suppress mode competition and achieved a broad oscillation. The obvious similarity of our technique to passive mode-locking of ultrafast lasers led us to adopt the term pair-wise mode locking for it.

# An Intense Continuous Source of Broadband Down-Converted Light

Avi Pe'er, Yaron Silberberg, Barak Dayan, and Asher A. Friesem
*Department of Physics of Complex Systems, Weizmann Institute of Science, Rehovot 76100, Israel*

The unique coherence and squeezing properties of high-power broadband down-converted light are very attractive for basic research in quantum optics as well as for applications. However, efficient sources did not exist until now, because of the need to stimulate the down-conversion process in a cavity, where mode competition narrows the spectrum drastically. Here we present the design and experimental realization of a low threshold optical parametric oscillator source that oscillates over the entire bandwidth allowed by phase matching. Two-photon absorption is exploited as a selective loss mechanism to suppress mode competition and achieve a broadband oscillation. In our source we perform pairwise mode-locking of many frequency pairs, in direct equivalence to passive mode-locking of ultrashort pulsed lasers.

PACS numbers: 42.65.Lm, 42.65.Re ,42.65.Yj, 42.50.Dv, 42.50.St, 42.55.-f

Quantum mechanically, the parametric down-conversion process can be viewed as the conversion of one energy quantum at a high frequency (a pump photon) into two lower energy quanta (signal photon and idler photon), while sum-frequency generation is the symmetric inverse process. Spontaneous parametric down-conversion is the key process for generation of entangled photon pairs for quantum optics experiments at low powers [1–4]. At high powers, the unique coherence and squeezing properties of down-converted light are simultaneously equivalent to those of ultrashort pulses in time and narrowband continuous lasers in frequency [5]. Thus, high power down-converted light is attractive to applications, such as optical spread spectrum communication [6] and "quantum" lithography [7], that exploit classical features, and suppression of spontaneous atomic decays [8] that exploit quantum mechanical properties. For these applications, an intense, efficient, low threshold source that generates broadband down-converted light is necessary. Since the non-linear interaction is weak, an oscillator with a high finesse cavity seems a reasonable choice. Unfortunately, mode competition in such a cavity narrows the spectrum of actual oscillation even though phase matching allows broad oscillation. Specifically, since all the down-converted mode pairs compete for the energy of the pump, the mode pair with the highest gain is dominant, while all other pairs are suppressed. Yet, the dominant mode in the cavity is always the one with the highest conversion efficiency, i.e. it has the best gain-loss relation. Thus, if the loss is selective, modes with lower gain may dominate because their loss is even lower. Consequently, if a broadband oscillation is desired, the cavity should be specially designed to include a selective loss mechanism that suppresses narrowband oscillations but does not affect broadband oscillations.

Before dealing with a specific cavity design, we review some of the special features of broadband down-converted light, specifically, the inherent phase and amplitude correlation between the signal $A_s(\omega)$ and the idler $A_i(\omega)$. Although each field in itself is incoherent white noise, the two fields are complex conjugates of each other [6, 9], as

$$A_s(\omega) = A_i^*(\omega_p - \omega), \tag{1}$$

where $\omega_p$ is the pump frequency. Such symmetry in the spectrum indicates that the envelope of the down-converted field in time $A(t)$ around the center frequency $\omega_p/2$ is real; i.e. $A(t)$ has only one quadrature in the complex plane [10], and leads to the special properties of two-photon absorption (TPA) and sum-frequency generation (SFG), when performed with down-converted light. In general, the TPA probability (or SFG intensity) $R$ at frequency $\Omega$ is given by

$$R(\Omega) \propto \left| \int d\omega A_s(\omega) A_i(\Omega - \omega) \right|^2. \tag{2}$$

When $\Omega \neq \omega_p$, the integrand contains uncorrelated signal-idler frequency pairs, leading to a negligible response, as expected when TPA is excited with incoherent white noise. Yet, when $\Omega = \omega_p$, the summation is over correlated frequency pairs so the random phase of the integrand cancels out, leading to a fully constructive interference, exactly as if the process was performed with coherent transform limited pulses. Accordingly

$$R(\Omega) \propto \left| \int d\omega \left| A_s(\omega) \right|^2 \right|^2. \tag{3}$$

The resulting spectrally narrow excitation (as narrow as the pump laser) can be controlled by manipulating the spectral phase relations within the down-converted spectrum. Such manipulations can be achieved with a small relative delay / dispersion [5].

We consider the cavity configuration of a doubly resonant optical parametric oscillator (OPO), schematically shown in Fig. 1, where both the signal and the idler resonate. In addition to the gain medium, the cavity also includes a two-photon loss medium, sandwiched between two opposite dispersions. The two-photon loss medium can be for example a two-photon absorber or sum-frequency generator (it's exact properties are discussed later on). The main thrust of our cavity design is to exploit the dispersion in the cavity to control the two-photon loss such that narrowband oscillations will suffer loss, while broadband oscillations will not and consequently become dominant. Specifically, the introduction of a small amount of dispersion changes the phase relations between the different frequency pairs of broadband oscillations and reduces drastically the efficiency of two-photon processes, due to destructive interferences between different pairs. Thus, broadband oscillations can avoid the two-photon loss while narrow oscillations cannot. In order for this dispersion not to affect down-conversion at the gain medium, the phase relations should be restored after passage through the loss medium by introducing the inverse dispersion.

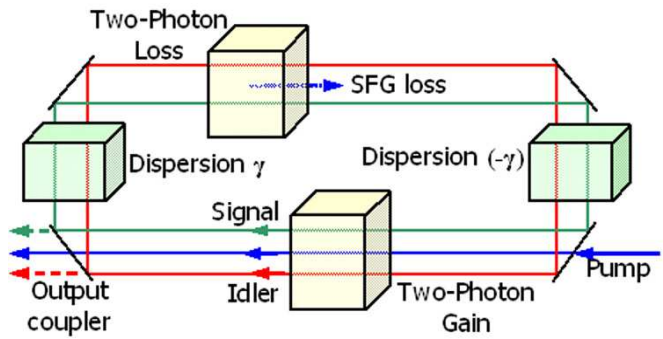

FIG. 1: A pairwise mode-locked OPO cavity configuration.

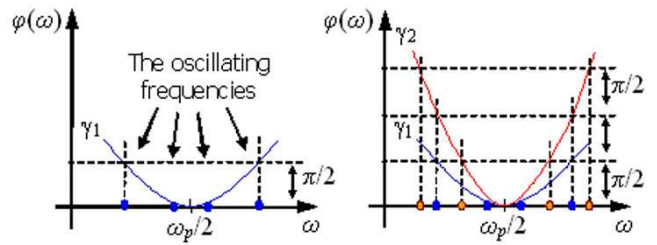

FIG. 2: The selection mechanism of the oscillating frequency pairs in the OPO cavity. (a) assuming a single two-photon loss medium at a dispersion value $\gamma_1$ and (b) assuming two media for two-photon loss at two different dispersion values ($\gamma_1,\gamma_2$).

Our approach for obtaining a broadband oscillation is inherently equivalent to passive mode-locking for obtaining ultrashort pulses. In both approaches a non-linear loss is inserted inside the cavity to enforce a broadband oscillation; in passive mode locking it is Kerr lensing or saturable absorption [11, 12], while in our approach it is a two-photon loss. In a mode-locked laser, ultrashort pulses are generated by locking the phases of all the single frequency modes to be equal, whereas in our approach ultrashort signal-idler temporal correlations are generated by locking the phases of all the frequency-pairs to be equal, namely "pairwise mode-locking". Accordingly, one can view the mode-locked laser as a broadband one-photon coherent source and a pairwise mode-locked OPO as a broadband two-photon coherent source.

Moreover, just as in ultrafast mode-locking, it is necessary to compensate for dispersion in the cavity, such that the total dispersion per pass is zero. A time domain reasoning for this is that efficient broadband down-conversion requires the signal-idler phase relations to be maintained after every pass in the cavity. A frequency domain explanation provides further clarification. Specifically, since $\omega_s+\omega_i=\omega_p$ the signal and the idler frequencies are exactly symmetric around the center frequency $\omega_p/2$. Yet, in a doubly resonant OPO, both the signal and the idler frequencies are from the frequency comb of the passive cavity. Consequently, in order to enable a broadband oscillation, the passive frequency comb should be as symmetric as possible. Since the comb is distorted by the total dispersion in the cavity, a symmetric comb requires that all even orders of dispersion be zero [13] (Odd orders of dispersion distort the comb symmetrically, not affecting the interaction). Since the total dispersion must be zero, the dispersion $\gamma$ introduced before the two-photon loss medium, must be compensated.

One of the most important properties of passively mode-locked lasers is that the frequency comb is exactly equally spaced, which is very appealing for precise measurements of optical frequencies [14, 15]. Although the frequency comb of our OPO cavity is not necessarily equally spaced, symmetry around the center frequency is guaranteed with essentially the same precision (or even better, due to squeezing).

Let us now examine how the spectrum emitted from the OPO is affected by the properties of the two-photon loss medium with the aid of Fig. 2. If the loss medium is non-dispersive, this spectrum is influenced only by the dispersion $\gamma$ between the gain and the loss media, which introduces a parabolic spectral phase $\exp\left(i\gamma\omega^2\right)$ onto the spectrum. We expect that the OPO oscillation will develop such that the two-photon loss is minimal [16]. Indeed, as shown in Fig. 2a, the two-photon loss can even be completely eliminated when the oscillation is composed of four frequencies. The phase of the outer frequency pair is shifted by $\pi$ ($\pi/2$ for every frequency) compared to the inner pair. As a result, the inner and the outer frequency pairs interfere destructively and there is no need for additional frequency pairs. However, for a truly broad oscillation, four frequencies are obviously not enough. To obtain more frequencies, we first consider having two loss media in the cavity at two different dispersion points $\gamma_1$ and $\gamma_2$ ($\gamma_2 \approx 2\gamma_1$). In this case, any four frequencies selected to minimize the loss at the first medium according to $\gamma_1$ will suffer high losses at the second medium according to $\gamma_2$. In order to minimize both losse simultaneously, the OPO must double the number of frequencies, as shown in Fig. 2b. Following this reasoning, three different dispersion values will lead to 16 frequencies and so on. Consequently, the ideal two-photon loss medium is a highly dispersive two-photon absorber, where dispersion is continuously accumulated along the loss medium.

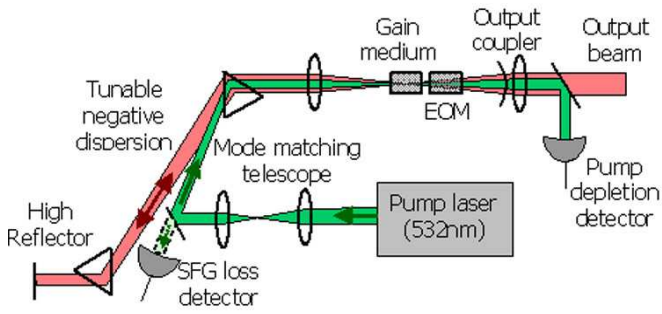

FIG. 3: Experimental OPO cavity configuration.

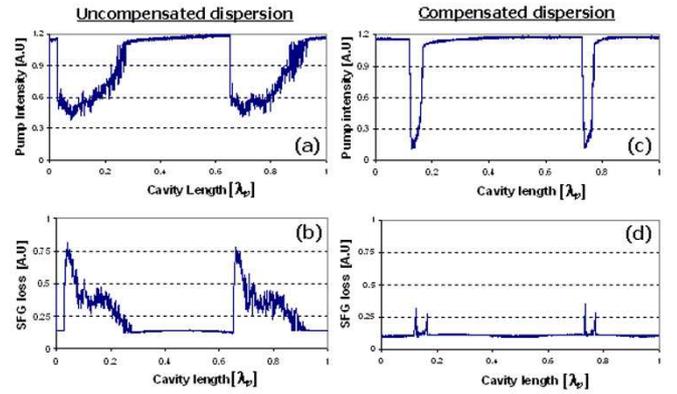

FIG. 4: Measurements of pump depletion and SFG loss as a function of the differential OPO cavity length over a range of one pump wavelength. (a) pump depletion when dispersion is not optimally compensated; (b) is the corresponding SFG loss. (c) pump depletion when dispersion is optimally compensated; (d) corresponding SFG loss.

Note that if we use SFG in a long dispersive crystal for the loss process the net result is just equivalent to having a single dispersion value (the middle value) since the SFG photons generated at the beginning of the crystal propagate through and coherently interfere with those generated at the end. Thus, in order to use SFG for the two-photon loss it is necessary to somehow incoherently "dispose" of the generated SFG photons. This can be achieved by absorbing the SFG photons within the crystal with some doping. Alternatively, it is possible to use walk-off between the down-converted and SFG beams. Specifically, in bi-refringent phase matching the polarizations of the down-converted and the SFG photons are orthogonal, so in most cases, the SFG beam is deflected by a walk-off angle. If the down-converted beam is tightly focused into the bi-refringent crystal, the SFG photons will leave the propagation path after a short distance, leading to a desired incoherence between the different dispersion values.

In order to demonstrate experimentally the principles of pairwise mode-locking we constructed the linear OPO cavity schematically depicted in Fig. 3. Gain in such a linear cavity OPO exists only when the down-converted light propagates through the medium along the direction of the pump. As a result, we exploited SFG that occurs during backward propagation as the two-photon loss. This SFG loss is normally a major limitation on the performance of a linear-cavity doubly resonant OPO; due to this loss, the conversion efficiency of the OPO cannot exceed 50% for narrowband oscillations [9, 16]. Since for broadband oscillations this limit no longer holds, it is expected that the measured conversion efficiency with our OPO would exceed 50%. In addition, since the SFG photons are emitted backward out of the OPO cavity, it is possible to isolate and detect them in order to measure the two-photon loss for both narrowband and broadband oscillations. A disadvantage of this configuration is that both the gain and the loss occur in the same crystal, so it is impossible to exploit the walk-off effect in order to obtain a continuum of dispersion values; effectively the configuration is comparable to one having a non-dispersive two-photon loss with a specific dispersion value. Accordingly, the oscillation spectrum is expected to contain four strong lobes instead of a truly broad spectrum.

As noted earlier, it is necessary to match the cavity frequency comb to the pump frequency in a doubly resonant OPO. This was achieved in our OPO with an electro-optic phase modulator (EOM) inside the cavity. An EOM was preferred over a piezo-controlled cavity mirror because it has a higher locking bandwidth, limited only by the bandwidth of the driving electronics. The positive dispersion in the cavity is provided mainly by the two crystals (EOM + gain medium) and the negative dispersion is introduced by a prism pair. The gain medium in our OPO cavity was a 12mm long periodically polled $KTiPO_4$ (PPKTP) crystal pumped at 532nm by a single-frequency doubled Nd:Yag laser (Verdi by Coherent). The bandwidth allowed by phase matching for this crystal is $\sim 50nm$ around 1064nm. The EOM was a 10mm long $RbTiPO_4$ (RTP) crystal and the prism pair was constructed from highly dispersive SF57 glass. All the intra-cavity elements were anti-reflection coated for 532nm and 1064nm. The output coupler transmission was 2%. Under these conditions a threshold of $\sim 0.3W$ for oscillation was measured.

The experimental measurements of the pump depletion and the SFG two-photon loss are presented in Fig. 4. All measurements were taken at a pump power of 1W. When the voltage applied to the EOM is varied, the length of the cavity is linearly scanned. Accordingly, Figs. 4a and 4c represent two different measurements of the pump intensity after passage through the cavity as a function of the differential cavity length in units of the pump wavelength. Obviously, the pump is strongly depleted whenever the cavity length meets the oscillation condition (twice in every scan). Figs. 4b and 4d are the corresponding measurements of the SFG two-photon loss, showing the loss that appears whenever an oscillation develops. Using the prism pair to tune the dispersion we performed these measurements twice: First, when a residual dispersion exists in the cavity (Figs. 4a and 4b) and then, when the dispersion is optimally compensated (Figs. 4c and 4d).

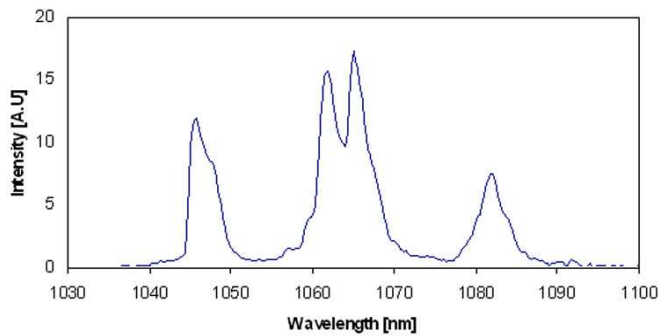

FIG. 5: A typical oscillation spectrum of the OPO configuration in Fig. 3.

When dispersion exists in the cavity, a broadband oscillation cannot develop since the oscillation condition is different for different frequency pairs. As a result, only a narrowband oscillation develops at any specific cavity length, and as expected, the measured SFG loss is high and the pump depletion low. However, when the dispersion is compensated, a broadband oscillation is allowed and indeed the SFG loss is negligible and the pump depletion very high. The results in Fig. 4b were obtained at the center of the pump beam with a small detector ($0.8mm^2$) and indicates depletion of $> 85\%$. The total depletion, measured with a large detector, was $> 60\%$ because of imperfect spatial overlap between the pump and the cavity mode. Such high pump depletion with negligible SFG two-photon loss, indicates that indeed the down-conversion efficiency exceeded 50%. However, due to high levels of intra-cavity loss and non optimal output coupling, the measured output power of the OPO was only $\sim 150mW$.

We then added a feedback loop to actively lock the OPO cavity length to the pump frequency in order to obtain a stable oscillation. We measured the emitted spectrum with a fast spectrometer (4ms integration time). A typical spectrum is shown in Fig. 5. The spectrum was unstable and changed on a time scale of $\sim 0.5s$, but was predominantly a four lobed spectrum as expected. The reason for the instability is that the four oscillating frequencies are not unique, so minute changes in the cavity, such as uncompensated acoustic vibrations, air currents, will cause the frequencies to spontaneously hop.

So far, we considered a doubly resonant OPO. However, the concept of pairwise mode-locking applies equally to singly resonant OPOs (only one of the signal/idler fields resonates). Indeed, assuming both signal and idler traverse the two-photon loss medium with opposite dispersions, the resulting spectrum will be similar. Although the oscillation threshold of a singly resonant OPO is higher, it's configuration can be much simpler, since it does not require active locking of the cavity to the pump frequency (the idler frequencies are not constrained by the cavity).

Although our analysis was purely classical, the potential non-classical properties of such an OPO are of great interest. Any OPO generates squeezed light, where the degree of squeezing is mostly limited by linear losses. Since our pairwise mode-locking involves only two-photon losses, it does not affect the signal-idler correlation, so it enables the generation of broadband squeezed light, which can be applicable to suppression of spontaneous emission [8] or to optical phase measurements at the Heisenberg limit [17, 18]. We developed an efficient OPO source of broadband down-converted light that has unique coherence and squeezing properties. We believe that the precise symmetry in the frequency comb of such a source will be advantageous for precision two-photon spectroscopy.

The authors wish to thank Prof. Nir Davidson for many helpful discussions. This research was partially supported by the Yshaayah Horowitz foundation.

# Chapter 6
# Performance Analysis
# for a Pair-wise Mode-Locked OPO

In order to analyze the dependence of the conversion efficiency on the bandwidth of oscillation for the pair-wise mode locked OPO (assuming steady state operation), I follow ref. 3 (Ch. 9, pp. 117-140), where an analysis of the threshold pump intensity and conversion efficiency is given for the case of monochromatic signal and idler. Specifically, I perform a similar analysis for broadband signal and idler with the proposed mode competition suppression scheme shown previously. For simplicity, I assume at the beginning just one non dispersive SFG two-photon loss medium in the cavity. Later I explain how the result is generalized for the case of several (or even a continuum of) dispersion values.

Assuming that the depletion of the pump is relatively low and that the gain per pass in the cavity is not very high, it is valid to assume that the intensity of the signal and the idler fields is constant throughout the cavity and to express the pump amplitude after the non-linear gain medium $A_p^+$ as:

$$A_p^+ = A_p^0 - l\chi \int d\omega A_s(\omega) A_i(\omega_p - \omega) \tag{1}$$

where $A_p^0$ is the pump amplitude entering the medium and I assumed the non-linear coupling constant $\chi$ to be independent of frequency, which is a reasonable assumption for frequencies close to the degeneracy point. Similarly, the amplitude of the SFG two-photon loss $A_{TPL}$ is

$$A_{TPL} = l\chi \int d\omega A_s(\omega) A_i(\omega_p - \omega) \exp[i\gamma\omega^2] \tag{2}$$

where $\gamma$ measures the dispersion accumulated during propagation from the gain medium to the two-photon loss medium. When other phase control mechanisms are used, Eq. (3) should be modified accordingly (without affecting the analysis).

Now, assuming the output coupler reflectivity to be equal for both the signal and the idler, the cavity conditions are symmetric, so the idler and the signal are complex conjugates:

$$A_i(\omega_p - \omega) = A_s^*(\omega).  \quad (3)$$

Incorporating Eq. (3) into Eqs. (1) and (2) yields

$$A_p^+ = A_p^0 - l\chi \int d\omega |A_s(\omega)|^2, \quad A_{TPL} = l\chi \int d\omega |A_s(\omega)|^2 \exp[i\gamma\omega^2]. \quad (4)$$

Defining the loss amplitude function $F(\gamma)$ as

$$F(\gamma) \equiv \int d\omega |A_s(\omega)|^2 \exp[i\gamma\omega^2]  \quad (5)$$

we obtain

$$A_p^+ = A_p^0 - l\chi F(0), \quad A_{TPL} = l\chi F(\gamma)  \quad (6)$$

Note that $F(0) = \int d\omega |A_s(\omega)|^2$ is proportional to the number of signal photons in the cavity, which is equal to the number of down converted photon pairs.

It is now possible to write an energy conservation equation, stating that in steady state the number of photons per second lost from the pump is equal to the number of signal-idler photon pairs leaving the cavity per second, as

$$|A_p^0|^2 - |A_p^+|^2 - |A_{TPL}|^2 = TF(0)  \quad (7)$$

where $T$ is the loss in the cavity, which is equal to the output coupler transmission in an ideal cavity. Substituting Eq. (6) into Eq. (7), and performing some algebra yields

$$\frac{TF(0)}{|A_p^0|^2} = \frac{1}{1 + |F(\gamma)/F(0)|^2} \left[ \frac{2T}{\chi l A_p^0} - \frac{T^2}{\chi^2 l^2 |A_p^0|^2} \right]  \quad (8)$$

The left hand side of Eq. (8) can be identified as the conversion efficiency $\eta$, since it is just the number of down converted signal-idler photon pairs leaving the cavity per second divided by the number of incident pump photons per second. Since I assume perfect phase matching, the pump field can be taken as real, and recalling the expression for the threshold pump intensity [3,4]

$$|A_{p-th}|^2 = \frac{T^2}{4\chi^2 l^2}  \quad (9)$$

we can rewrite Eq. (8) as

$$\eta = \frac{4}{1 + |F(\gamma)/F(0)|^2} \left[ \left|\frac{A_{p-th}}{A_p^0}\right| - \left|\frac{A_{p-th}}{A_p^0}\right|^2 \right]  \quad (10)$$

Defining $N$ as the ratio between the actual pump intensity ($I_p$) and the threshold pump intensity ($I_{th}$),

$$N \equiv \frac{I_p}{I_{th}} = \left|\frac{A_p^0}{A_{p-th}}\right|^2 \tag{11}$$

we obtain:

$$\eta = \frac{2}{1+|F(\gamma)/F(0)|^2}\left(\sqrt{N}-1\right)\frac{2}{N}. \tag{12}$$

Equation (12) indicates that assuming equal thresholds (equal *N*), the dominant oscillation will be the one that minimizes the two-photon "tax" $|F(\gamma)|^2$ for any pumping power (for any *N*); i.e. a broad oscillation. As explained in the article, several (preferably a continuum of) independent loss media at different dispersion values are required for a stable broad oscillation. In this case, the dominant oscillation would be the one that minimizes the total "tax":

$$\sum_m |F(\gamma_n)|^2 \to \int d\gamma |F(\gamma)|^2, \tag{13}$$

where the integral expression stands for the continuum limit.

Comparing the two limiting possibilities of a very narrow oscillation, where the two-photon "tax" is essentially independent of *γ*, and a very broad oscillation, where the two-photon loss tends to zero for a large enough dispersion, it is seen that the improvement in conversion efficiency approaches a factor of two, which is a considerable improvement indicating why a broad oscillation would be highly favored. In practice, it is expected that a broad oscillation will have a higher threshold. Thus, when the pump power is low, narrow oscillations will dominate. But, as the pump power increases high above threshold, the situation becomes more and more favorable for the broader oscillations. The conversion efficiency as a function of *N* is given in figure 1 for three cases – (1) very narrow oscillation (dashed curve), (2) ideal broad oscillation (dotted curve) and (3) practical broad oscillation (solid curve). Note that for very broadband oscillations, where the two-photon loss becomes negligible, the conversion efficiency can approach 100% around *N* = 4. This, of course, is most desirable for many applications.

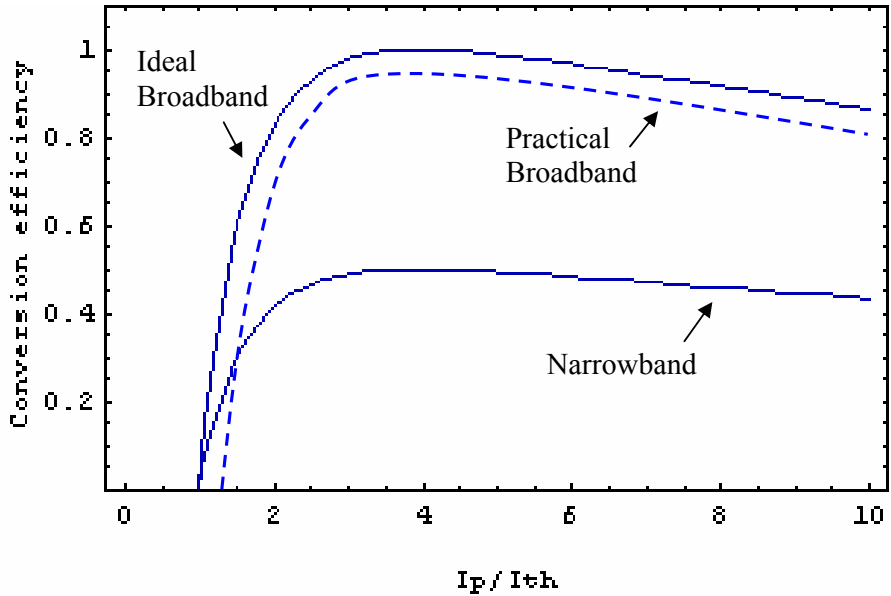

Figure 1: Calculated conversion efficiency as a function of $N = I_p/I_{th}$ for a narrow oscillation (dashed), ideal broadband oscillation (dotted) and practical broadband oscillation (full line). $I_{th}$ is taken to be the minimum threshold intensity among all possible oscillations.

Figure 1 indicates that, in order to obtain broadband oscillations, it is desirable that the threshold for broad oscillations will be equal to the minimum threshold of the narrowband oscillation. In other words, since broadband oscillation can be decomposed into many signal-idler frequency pairs, it is desired that all these pairs will have the same threshold – i.e. that the threshold intensity will be independent of wavelength. This requirement can be met to a high degree as shown in figure 2, where a calculation of threshold intensity as a function of wavelength is given for the two cases of broad phase matching discussed earlier. The calculated graph for a PPKTP crystal pumped at 532nm is shown in figure 2a and for a BBO crystal pumped at 728nm in figure 2b. As evident, the threshold intensity is constant up to 15% over the entire phase matching bandwidth.

The performance analysis given here serves as a basic theoretical model for our pair-wise mode-locking approach. Combined with the experiments described in the previous chapter this analysis gives a sound proof that a low threshold OPO source emitting a broad spectrum of down converted light is indeed feasible.

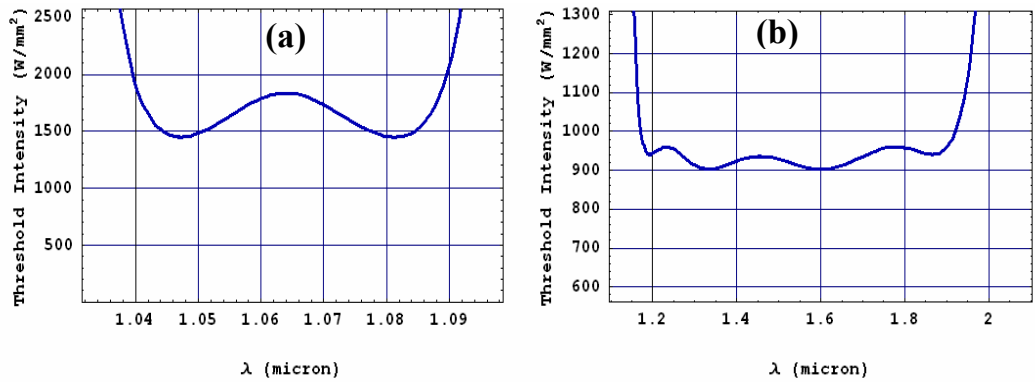

Figure 2: Threshold pump intensity ($I_{th}$) as function of signal wavelength. **(a)** For a 10mm long periodically polled KTP crystal with 4% loss in the cavity pumped by 532nm in a broad phase matching configuration. **(b)** For a 14mm long BBO crystal with 1% loss in the cavity pumped by 728nm in an ultra broad, zero dispersion, phase matching configuration.

# Chapter 7
# Nonlinear Interactions and Temporal Shaping with Broadband Entangled Photons

We experimentally investigated the properties of down-converted light at low power, where it can be considered as composed of a stream of separate entangled photon pairs and it's non classical nature is most prominent. Even though the average occupation of each signal or idler frequency mode at such low light powers is less than unity, twin modes are either occupied or unoccupied together. This entanglement between frequency mode pairs strongly affects non-linear interactions. Specifically, SFG and TPA are predicted to depend linearly on the flux of incoming photons [49-56]. This linear dependence at low power is a clear indication of the discrete quantum nature of the light field.

In order to overcome the low efficiency of non-linear interactions at the single photon level, we generated in our experiment a flux of about $10^{12}$ entangled-pairs per second (~0.3μW). The key element in achieving such a high photon-flux while maintaining its non-classical properties, is generating the photons as broadband as possible, since the coherent addition of many entangled frequency modes can enhance the effect dramatically just as it does in high power. We demonstrated the non-classical linear flux dependence of the non-linear SFG process at the low power regime. We then utilized the SFG process as an ultrafast coincidence detector to directly measure the two-photon correlation and to coherently control it.

# Nonlinear Interactions with an Ultrahigh Flux of Broadband Entangled Photons

Barak Dayan, Avi Pe'er, Asher A. Friesem, and Yaron Silberberg

*Department of Physics of Complex Systems, Weizmann Institute of Science, Rehovot 76100, Israel*

We experimentally demonstrate sum-frequency generation with entangled photon pairs, generating as many as 40 000 photons per second, visible even to the naked eye. The nonclassical nature of the interaction is exhibited by a linear intensity dependence of the nonlinear process. The key element in our scheme is the generation of an ultrahigh flux of entangled photons while maintaining their nonclassical properties. This is made possible by generating the down-converted photons as broadband as possible, orders of magnitude wider than the pump. This approach can be applied to other nonlinear interactions, and may become useful for various quantum-measurement tasks.

Entangled photons play a dominant role in quantum communication science due to their inherent nonclassical correlations [1–5]. Considerable effort has been invested towards achieving nonlinear interactions with entangled photons, since such interactions are significant for all-optical quantum computation and quantum metrology [6–9]. Most works have focused on enhancing the nonlinear photon-photon coupling in order to achieve conditional phase shifts with single photons. Several strategies aimed at increasing the low efficiencies of nonlinear interactions with single photons (typically below $10^{-10}$) were developed. One such strategy is increasing the nonlinearity through strong photon-atom coupling in a high-finesse cavity [9–11]; another uses mixing with a strong pump that is coherent with the interacting photons [12]. Other proposed schemes rely on enhanced nonlinearities obtained through electromagnetically induced transparency [13–15]. Resonantly enhanced two-photon absorption was performed with narrowband down-converted light at power levels that approached the entangled-photons regime, demonstrating a nonclassical departure from quadratic intensity dependence [16]. At lower power levels, where down-converted light can be considered as composed of separated photon pairs, a completely linear intensity dependence of two-photon absorption and sum-frequency generation (SFG) is predicted [17–20]. Such a linear dependence can be shown to break the Cauchy-Schwartz inequality for classical fields [21].

Here we propose an alternative approach towards achieving nonlinear interactions with entangled photons, which does not focus on increasing the nonlinear coupling, but rather on obtaining an ultrahigh flux of entangled photon pairs. In our setup we generate a flux of about $10^{12}$ entangled pairs per second, a flux that corresponds to a classically high power level of 0.3 $\mu$W. This flux is orders of magnitude greater than is typically utilized in quantum-optics experiments, which are usually limited to electronic detection rates.

The key element in achieving such a high photon flux while maintaining its nonclassical properties is generating the entangled pairs as broadband as possible. The physical reason for this is that the arrival times of the two photons are correlated to within a time scale that is inversely proportional to their bandwidth [1]. Thus, the maximal flux $\Phi_{\max}$ of down-converted photons that can still be considered as composed of distinct photon pairs scales linearly with the down-converted bandwidth $\Delta_{\rm DC}$ [18,22]:

$$\Phi_{\max} \approx \Delta_{\rm DC}. \qquad (1)$$

This maximal flux, which corresponds to a mean spectral photon density of $n = 1$ (one photon per spectral mode), is the crossover point between the classical high power regime and the nonclassical low power regime.

A complete quantum-mechanical analysis [22] shows that the rate of SFG events with correlated pairs (i.e., photons generated from the same pump mode) indeed includes a term that depends linearly on the intensity:

$$R_c \propto \Delta_{\rm DC}(n^2 + n). \qquad (2)$$

The dependence on $\Delta_{\rm DC}$ indicates that the probability for up-conversion is inversely proportional to the temporal separation between the photons. This dependence on the bandwidth is not unique to entangled photons; the rate of SFG induced by classical pulses demonstrates the same dependence on their bandwidth, for the same reason. It is only the linear dependence on $n$ in Eq. (2) that is unique to entangled photons. Note that all up-converted correlated pairs produce SFG photons back at the wavelength of the pump. Accordingly, in Eq. (2) it was assumed that the pump bandwidth $\delta_p$ is completely included within the available up-converted bandwidth $\delta_{\rm UC}$ (i.e., the bandwidth that is phase matched for up-conversion in the nonlinear crystal).

Apart from the SFG events that are generated by correlated pairs there is always a classical background noise due to "accidental" SFG of uncorrelated photons. Unlike correlated pairs, uncorrelated pairs can be up-converted to any wavelength in the range of $2\Delta_{\rm DC}$ around the pump frequency. The actual rate $R_{u.c.}$ of such SFG events is therefore proportional to the available up-converted bandwidth $\delta_{UC}$ [22]:

$$R_{u.c.} \propto \delta_{UC} n^2. \quad (3)$$

Like in many other detection schemes, this expression simply measures the background noise that falls within the spectrum of the "receiver."

To clarify the role of entanglement in the SFG process, let us first consider the narrowband case, where only one pair of signal and idler modes is involved in the down-conversion process. Assuming a low down-conversion efficiency that yields a mean spectral photon density of $n \ll 1$, the state $|\psi\rangle$ of the down-converted light can be described by

$$|\psi\rangle \approx M|0\rangle + \sqrt{n}|1\rangle_s|1\rangle_i, \quad (4)$$

with $M \sim 1$. The subscripts $s, i$ denote the signal and idler modes, whose frequencies sum to the pump frequency. Note that the spectral width of the modes can be defined as the spectral width of the pump mode $\delta_p$ (this means we quantize the fields according to the longest relevant time scale, which is the coherence time of the pump, and so the number of photons is defined over a time period of $\delta_p^{-1}$ [23]). The state $|\psi\rangle$ clearly describes an entanglement between the signal and idler modes, which are both either in the vacuum state $|0\rangle$ or in the one-photon Fock state $|1\rangle$. It is quite straightforward to show that this entanglement is the source for the linear term in Eq. (2). When the down-converted bandwidth is significantly larger than the pump bandwidth, time and energy entanglement is created between the photons. This actually means that the photon pairs can be generated in $N = \Delta_{DC}/\delta_p$ different entangled mode pairs:

$$|\psi\rangle \approx M|0\rangle + \sum_{j=1}^{N} \sqrt{n} |1\rangle_{s_j}|1\rangle_{i_j}. \quad (5)$$

Although the phase of each signal or idler mode separately is inherently uncertain, the excitations of these mode pairs are mutually coherent, all having a combined phase that is shifted by $\pi/2$ from the pump phase. Thus, the SFG probability amplitudes induced by all these pairs add coherently, resulting in an amplification by $N^2$ of the correlated SFG rate, compared to the narrowband case. Intuitively speaking, one factor of $N$ comes from having $N$ more photon pairs, and the second one comes from the fact that the temporal separation between the photons of each pair is smaller by a factor of $N$. On the other hand, The SFG induced by up-conversion of uncorrelated pairs is summed incoherently, resulting in an enhancement only by a factor of $N$. Indeed, by dividing Eqs. (2) and (3), we obtain that the ratio between the correlated and uncorrelated rates is

$$\frac{R_c}{R_{u.c.}} \approx \frac{\Delta_{DC}}{\delta_{UC}}\left(\frac{n+1}{n}\right) \leq N\left(\frac{n+1}{n}\right), \quad (6)$$

where the inequality results from the condition of $\delta_{UC} \geq \delta_p$. As is evident, this ratio is decreased when the up-converted bandwidth is larger than the pump bandwidth; hence the importance of performing a narrowband SFG. Note that this gain of the correlated process over the uncorrelated one holds at high powers as well, and affects not only SFG, but any other nonlinear mixing between the signal and idler fields (e.g., two-photon absorption [24,25]).

It is important to point out that while the gain described in Eq. (6) results from correlations, such correlations do not necessarily require entanglement. Coherent correlations between two broadband fields are the basic principle of all spread-spectrum communication schemes [26,27]. In fact, without the nonclassical linear term, Eqs. (2), (3), and (6) are identical to the equations for the signal and noise in spread-spectrum communication performed with shaped classical pulses [28]; one has only to replace $\Delta_{DC}$ with the bandwidth of the pulses and $\delta_p$ with the spectral resolution of the phase modulations performed on the pulses. However, while coherent pulses can be easily shaped to exhibit spectral phase and amplitude correlations, such correlations between Fock states are inherently nonclassical and imply entanglement. This entanglement is manifested in the fact that the gain in Eq. (6) is stronger by a factor of $(n+1)/n$ than is classically achievable, due to the additional, nonclassical linear term in the correlated SFG process. At high power levels the linear term becomes negligible and the correlations between the fields are almost identical to the correlations that can be obtained by shaping classical pulses. This is not to say that the increased precision of the correlations (i.e., the squeezing) vanishes at high powers, yet its influence on the SFG process becomes negligible. All previous experiments that involved SFG with down-converted light [29–31] were performed at power levels that greatly exceeded $\Phi_{max}$, where the intensity dependence was completely quadratic. The correlation effects observed in these experiments are well described within the classical framework, and are identical to those demonstrated with shaped classical pulses [28,32].

To conclude this discussion we summarize that all the nonclassical properties exhibited in SFG of down-converted light result from the part of the process that has a linear intensity dependence. This part exists due to the entanglement between signal and idler modes, and becomes negligible in high powers. Although time and energy entanglement is not required for the nonclassical behavior, it amplifies its effect with respect to the classical, uncorrelated SFG process.

Despite the fact that the SFG process exhibits a linear intensity dependence at $n \ll 1$, it is nonetheless a two-photon process. Thus, a random loss of either a signal or an idler photon is equivalent to the loss of the entire pair, and will lead to a ''classical'' quadratic reduction of the SFG signal. We expect therefore two different intensity dependencies of the SFG process: linear dependence, when the two-photon production rate is changed; quadratic dependence, when the entangled photon flux is attenuated by any linear-optics mechanism.

In our experiment we used a single-frequency ($\delta_p \approx$ 5 MHz around 532 nm) doubled Nd:YAG laser to pump a 12 mm long periodically poled KTiOPO$_4$ (PPKTP) crystal, generating infrared (IR) entangled photons with a broad bandwidth of $\Delta_{DC} = 31$ nm around 1064 nm. According to Eq. (1), this bandwidth implies a crossover flux of $\Phi_{max} = 8.2 \times 10^{12}$ photons per second, i.e., about 1.5 $\mu$W. A similar PPKTP crystal was used for the SFG process. The phase-matching conditions for the up-conversion process in this crystal were tuned to obtain an up-converted bandwidth of $\delta_{UC} \approx 100$ GHz around 532 nm. According to Eq. (6), these bandwidths ensure that the correlated SFG process dominates at any power.

The experimental setup is depicted in Fig. 1. Basically, entangled photons down-converted in one crystal were up-converted in the other crystal to produce the SFG photons. The entire layout was designed to maximize the interaction and collection efficiencies. The noncritical phase matching of the PPKTP crystals eliminates the ''walk-off'' between the down-converted photons and the pump, and allows optimal focusing [33] of the pump on the first crystal and of the entangled photons on the second one. The optimal focusing and the high nonlinear coefficient of the crystals ($\sim 9$ pm/V) yielded high nonlinear interaction efficiencies of up to $10^{-7}$. Both crystals were temperature stabilized to control their phase-matching properties.

For the SFG photons to be detected distinctly, any residue of the pump had to be filtered out from the down-converted photons by a factor of at least $10^{-18}$. We chose to do so with a set of four dispersion prisms, designed for refraction at the appropriate Brewster angle. This arrangement had two major advantages over schemes which rely on harmonic filters for filtering out the pump. First, Brewster-angle prisms enabled a very low loss of down-converted photons. Second, this layout enabled a tunable compensation of dispersion (mainly from the crystals), thereby avoiding a significant reduction of the bandwidth effective for the SFG process. The prisms were made of highly dispersive SF6 glass, in order to minimize the dimensions of our setup. The entangled-photon beam was filtered out from the SFG photons by a harmonic-separator mirror, and its power was measured by a sensitive InGaAs detector. The SFG photons were further filtered by line filters for 532 nm and counted with a single-photon counting module (SPCM) from EG&G (model AQR-15).

In order to verify that any photon detected by the SPCM was the result of the SFG process, and not a residue of the pump, we destroyed the phase matching in the second crystal by changing its temperature, and observed the SPCM's count drop to its dark count ($\sim 50$ s$^{-1}$), even with the pump running at full power (5 W). This dark-count level was subtracted from all the subsequent measurements, which were performed with integration times of 2–5 s, and with pump powers that did not exceed 2.5 W.

We measured the power dependence of the SFG process in two ways: one, by attenuating the entangled-photon beam with optical attenuators; the other by reducing the power of the pump laser. Figure 2 depicts the results of these two measurements. As expected, attenuation resulted in a classical, purely quadratic decrease of the SFG counts. However, reducing the entangled-pair flux by decreasing the pump power resulted in a nearly linear decrease of the

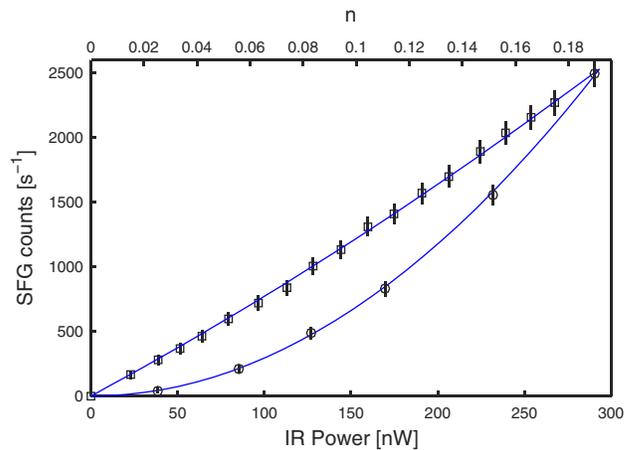

FIG. 2 (color online). Power dependence of SFG with entangled photons. When the down-converted IR power is reduced by optical attenuators, the SFG rate drops quadratically: circles, experimental; line, quadratic fit, exactly like classical light sources. However, when the IR power is reduced by decreasing the pump-laser power, thus reducing the photon-pair production rate, the SFG rate decreases in a close-to-linear manner, as quantum mechanically predicted: squares, experimental; line, calculated $\alpha(n + n^2)$, with $\alpha$ chosen to fit. The log-log slope of these measurements varies from 1.01 at the lowest powers to 1.14 at the highest powers.

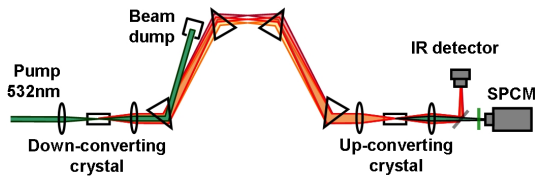

FIG. 1 (color online). Experimental layout. Entangled photons generated by down-conversion of a pump laser in one crystal are imaged through a set of four dispersion prisms onto a second crystal to generate the SFG photons. The entangled-photon beam is separated from the SFG photons by a harmonic-separator mirror and its power is measured by an InGaAs detector. The SFG photons are further filtered by 532 nm line filters and are counted with a single-photon counting module.

SFG signal. The slight deviation from linearity at high powers is in complete agreement with our calculations, validating that the entire measurement was performed in the regime of $0 < n < 0.185$, i.e., well below the crossover flux. The maximal SFG count in this experiment was about $2500~\text{s}^{-1}$. Taking into account that the collection efficiency of the SFG photons (limited by the transmission of the filters and the detector's efficiency) was about 6%, the number of generated SFG photons actually reached $40\,000~\text{s}^{-1}$, a flux that was visible even to the naked eye.

While our approach still suffers from a low nonlinear efficiency, and thus is not readily applicable for quantum computation, it nonetheless introduces new capabilities for quantum metrology, since it reveals the nonclassical correlations between entangled photons with fidelity that exceeds that of electronic coincidences. Specifically, the SFG process simultaneously measures both the time difference and the energy sum of the photons, without measuring their individual energies or times of arrival. The large gain of $N = \Delta_{\text{DC}}/\delta_{\text{UC}}$ over the uncorrelated SFG rate depends directly on this effect, which implies that narrowband SFG of time and energy entangled photons automatically rejects coincidences of nonentangled photons. This valuable property is unattainable with electronic coincidence detection, since electronic photodetectors detect the actual arrival time of each photon separately, and therefore must be as broadband as the photons themselves.

Our work points at the possibility of maintaining the nonclassical properties of entangled photons even at classically high powers by utilizing broadband, continuous down-conversion [34]. We believe this ability to perform nonlinear interactions with ultrahigh fluxes of broadband entangled photons holds promise in quantum-measurement science, in particular, for phase measurements at the Heisenberg limit [35].

This research was supported by the Yeshaya Horowitz Association.

# Temporal Shaping of Entangled Photons

Avi Pe'er, Barak Dayan, Asher A. Friesem, and Yaron Silberberg

*Department of Physics of Complex Systems, Weizmann Institute of Science, Rehovot 76100, Israel*
(Received 18 November 2004; published 22 February 2005)

We experimentally demonstrate shaping of the two-photon wave function of entangled-photon pairs, utilizing coherent pulse-shaping techniques. By performing spectral-phase manipulations we tailor the second-order correlation function of the photons exactly like a coherent ultrashort pulse. To observe the shaping we perform sum-frequency generation with an ultrahigh flux of entangled photons. At the appropriate conditions, sum-frequency generation performs as a coincidence detector with an ultrashort response time ($\sim 100$ fs), enabling a direct observation of the two-photon wave function. This property also enables us to demonstrate background-free, high-visibility two-photon interference oscillations.

Entangled photons generated by parametric down-conversion can exhibit nonclassical correlations between various physical properties, such as polarization, momentum, and energy, and have been the primary tool in quantum communication science [1–5]. While the polarization state of polarization-entangled photons has been readily manipulated in various quantum-optics experiments (see, for example, [3–5]), no control over their temporal properties has been demonstrated to date, except the prolongation of their correlation time by spectral filtering with narrowband filters [6–8].

Here we demonstrate that the two-photon wave function of entangled photons can be shaped by spectral-phase manipulations exactly like classical pulses. The physical reason for this lies in the coherent spectrum of the two-photon state, which governs the temporal correlations between the photons in the very same way that the spectrum of a coherent pulse determines its shape.

Let us consider the two-photon state $|\Psi\rangle$ created by type-I, broadband down-conversion of a narrowband pump [9]:

$$|\Psi\rangle \propto M|0\rangle + \int d\omega g(\omega)|\omega\rangle_s|\omega_p - \omega\rangle_i, \qquad (1)$$

where $M \sim 1$, and the subscripts $s, i, p$ denote the signal, idler, and pump modes, respectively. $g(\omega)$ is determined by the nonlinear coupling and the phase-matching conditions in the down-converting crystal and exhibits a constant spectral phase for the case of perfect phase matching. The probability for the joint detection of signal and idler photons at times $t_s, t_i$, respectively, is proportional to the second-order correlation function $G^{(2)}(t_s - t_i)$. At low photon fluxes, the two-photon wave function $\psi(t_s, t_i)$ represents the probability amplitude for such a detection (i.e., $G^{(2)}(t_s - t_i) = |\psi(t_s, t_i)|^2$), and is defined by

$$\psi(t_s, t_i) = \langle 0|E_s^+(t_s)E_i^+(t_i)|\Psi\rangle. \qquad (2)$$

Introducing spectral filters $\Theta_s(\omega), \Theta_i(\omega)$ to the signal and idler modes, and assuming a Gaussian spectrum with a bandwidth $\delta_p$ for the pump, the two-photon wave function can be approximated by [10,11]

$$\psi(t_s, t_i) \propto e^{-(1/32)\delta_p^2(t_s+t_i)^2} F(t_s - t_i), \qquad (3)$$

where $F(t)$ is the inverse Fourier transform of $g(\omega)\Theta_s(\omega)\Theta_i(\omega_p - \omega)$ as a function of $\omega$. In other words, the two-photon wave function behaves like a coherent pulse with a spectrum $g(\omega)$, which was shaped by a spectral filter $\Theta(\omega) = \Theta_s(\omega)\Theta_i(\omega_p - \omega)$. This shaping can be nonlocal, since both the filtering and the detection of the signal and the idler photons can be performed at separate locations. Note that, similarly, the two-photon wave function of momentum entangled photons (see [12]) could be spatially shaped by phase manipulations in the momentum space. In collinear down-conversion (as in our case), it is convenient to define the higher (lower) energy photon as the signal (idler), or vice versa. In such a case $g(\omega)$ is symmetric about $\omega_p/2$.

Although our discussion so far dealt with continuous down-conversion, shaping of the two-photon wave function is possible with pulsed down-conversion as well. In this case, the signal and the idler are both pulses with a constant (though undefined) spectral phase, like the pump. Thus, obviously, each can be shaped independently by spectral-phase filters, leading to the same result as in the continuous case [13].

The most direct way to observe the simultaneous arrival of two photons is to detect a nonlinear photon-photon interaction between them. Because of the extremely low efficiencies of nonlinear interactions with entangled photons, the detection of their inherent nonclassical correlations has been limited so far to photoelectric coincidences. Nonetheless, as was recently demonstrated [14], it is possible to generate surprisingly high fluxes of entangled photons without losing their nonclassical properties. Since the photons of each photon pair are temporally correlated to within a time scale that is inversely proportional to their bandwidth $\Delta_{DC}$, the maximal flux $\Phi_{max}$ at which down-converted light can still be considered as composed of separated photon pairs is

$$\Phi_{max} \approx \Delta_{DC}. \qquad (4)$$

This flux corresponds to a mean spectral photon density of $n = 1$. For broadband down-conversion, $\Phi_{max}$ can exceed $10^{13}$ photons per second, which corresponds to classically high-power levels of about 2 $\mu$W. Such ultrahigh fluxes of broadband entangled photons enable the demonstration of nonlinear interactions even with interaction efficiencies of the order of $10^{-9}$, which can be achieved without special enhancement mechanisms (see [14]).

In this work we introduce the use of sum-frequency generation (SFG) as an ultrafast coincidence detector for photon fluxes that are below $\Phi_{max}$. The sensitivity of the SFG process to a relative delay between the photons is inversely proportional to the interaction's bandwidth, which for broadband phase matching can be of the order of $\sim$100 fs. Such response times are completely unattainable with current electronic technologies and allow a direct observation of the two-photon wave function.

Our experimental setup followed that of [14] and is depicted in Fig. 1. Basically, entangled photons generated by down-conversion of a continuous pump laser ($\delta_p \approx$ 5 MHz around 532 nm) in one crystal were imaged through a set of four Brewster-angle dispersion prisms onto a second crystal to generate the SFG photons. This arrangement enabled a complete filtering out of the pump with very low loss of down-converted photons; it also enabled a tuned compensation of dispersion (mainly of the crystals), thereby avoiding a significant reduction of the bandwidth that was effective for the SFG process. Finally, the spectral separation induced by the prisms made it possible for us to turn the entire optical layout between the crystals into a pulse shaper. Specifically, by using the lens after the first crystal to focus the beam at the midway between the second and third prisms, we created there a spectral Fourier plane. Then, by placing a computer-controlled SLM in that plane, we were able to introduce arbitrary phase shifts to different spectral components of the entangled photons [15]. The nonlinear crystals were periodically poled KTiOPO$_4$ (PPKTP) crystals, which were temperature controlled to obtain broadband phase matching of $\Delta_{DC} \approx$ 31 nm around 1064 nm. This bandwidth implies $\Phi_{max} = 8.2 \times 10^{12}$ s$^{-1}$, which corresponds to the classically high-power level of about 1.5 $\mu$W. Our measurements were performed at power levels of about 0.25 $\mu$W ($n \approx 0.16$).

To observe the two-photon wave function we introduced a relative delay between the signal and idler photons of each pair, by using the SLM to apply linear spectral-phase functions with opposite slopes to the wavelengths below the degeneracy point of 1064 nm (signal) and above it (idler), as depicted in Fig. 2(a). As described earlier, in

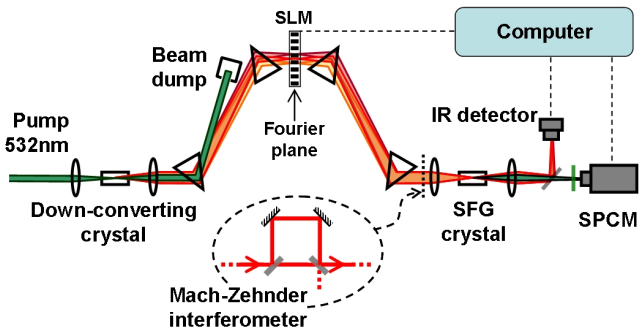

FIG. 1 (color online). Experimental layout. Entangled photons down-converted in one crystal are up-converted in the other crystal to produce the SFG photons. The symmetric imaging arrangement was designed to have a focus in the midway between the second and third prisms, creating there a spectral Fourier plane. A computer-controlled spatial light modulator (SLM) located at this plane was used to perform spectral-phase manipulations on the entangled photons. The SFG photons were counted using a single-photon counting module (SPCM-AQR-15 from *EG&G*), with a dark count of $\sim$50 s$^{-1}$. This dark count was subtracted from all our measurements, which were performed with integration times of 5–10 s. The entangled-photon beam was filtered out from the SFG photons by a harmonic-separator mirror and filters, and its power was measured by a sensitive InGaAs detector. In order to demonstrate two-photon interference oscillations, a Mach-Zehnder interferometer was placed between the last prism and the SFG crystal.

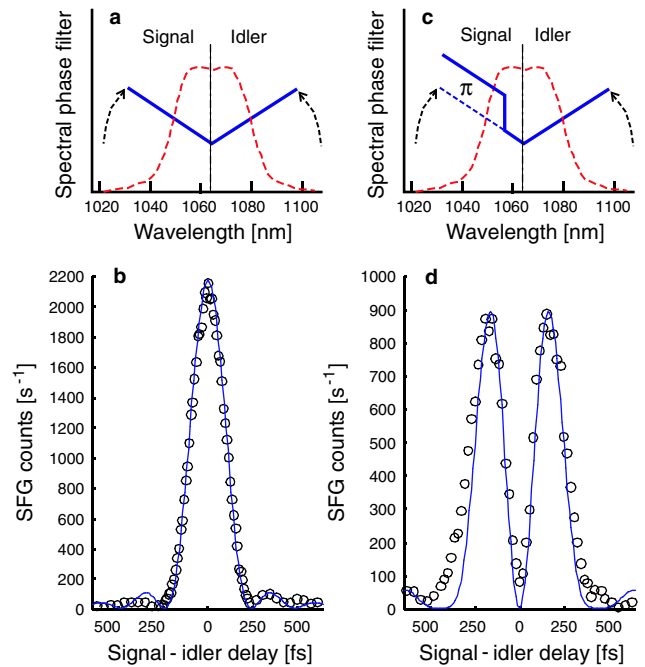

FIG. 2 (color online). Direct observation and shaping of the two-photon wave function. The relative delay between the signal and idler photons was induced by varying the slopes of the spectral-phase filter described in (a). (b) The SFG counts (circles) and the calculated second-order correlation function $G^{(2)}(t_s - t_i) = |\psi(t_s, t_i)|^2$ (line) as a function of the signal-idler delay. Shaping of the two-photon wave function was performed by adding a phase step function in the middle of the signal spectrum, as described in (c). The SFG measurements of the shaped wave function are represented in (d).

this case of continuous, broadband down-conversion, a delay has no separate effect on either the signal or the idler photons. Nevertheless, as depicted in Fig. 2(b), the response of the SFG rate to the relative delay between the signal and the idler photons directly reflects the two-photon wave function. In order to demonstrate our ability to coherently shape the two-photon wave function we applied a phase step function with amplitude of $\pi$ at the middle of the signal spectrum [roughly at 1057 nm; see Fig. 2(c)]. With coherent pulses, this has the effect of splitting a pulse in time to two lobes. Figure 2(d) depicts exactly such a behavior of the two-photon wave function. We note that these two lobes have opposite signs, which means there is a phase shift between the events where the signal photon arrives before the idler and the events where it arrives after it. This phase shift, although undetectable by any measurement performed on the signal (or idler) photon itself, can be detected by interfering the SFG field with the pump laser.

It is important to understand under which conditions SFG can be considered a good measure for photon coincidences. The phase-matching conditions in the nonlinear crystal dictate two relevant bandwidths (or three, in the nondegenerate case) for the process; one is the phase-matched bandwidth of the input (low-frequency) fields $\Delta_{LF}$, and the other is the phase-matched bandwidth for the up-converted field $\delta_{UC}$. Loosely speaking, for a SFG event to occur, three conditions must be satisfied. First, the spectrum of the input photons must lie within $\Delta_{LF}$. Second, the spectrum of the up-converted photon must lie within $\delta_{UC}$. Third, the incoming photons must overlap temporally to within $\Delta_{LF}^{-1}$. Obviously, only the third condition is equivalent to a coincidence detection. Therefore, SFG can be considered a coincidence detector only if the entire bandwidth $\Delta$ of the photons can be up-converted to a phase-matched wavelength, i.e., only if

$$\Delta_{LF} \gtrsim \Delta, \qquad \delta_{UC} \gtrsim 2\Delta. \quad (5)$$

Typically, the second condition in Eq. (5) does not hold, since in most cases $\delta_{UC} \ll \Delta_{LF}, \Delta$, especially when long crystals are utilized to increase the interaction efficiency. However, this condition can be circumvented with energy entangled photons. At low down-converted photon fluxes most of the SFG events are the result of the up-conversion of entangled pairs. Specifically, the rate of SFG events with spontaneously down-converted light can be approximated by

$$R \propto \Delta_{DC}(\delta_{UC} n^2 + \Delta_{DC} n^2 + \Delta_{DC} n). \quad (6)$$

As described in [14], the last term, which exhibits a nonclassical linear intensity dependence and therefore dominates at $n \ll 1$, is the one that results from SFG of entangled pairs [16]. Since the sum energy of such pairs is always equal to the energy of the pump photons, for entangled-photon fluxes below $\Phi_{max}$ the second condition of Eq. (5) can practically be replaced by

$$\delta_{UC} > \delta_p. \quad (7)$$

This condition is easily met for continuous down-conversion, where the pump is typically very narrowband, as was in our case, where $\Delta_{LF} = \Delta = \Delta_{DC}$ (since identical crystals were used for the down- and up-conversions), and $\delta_{UC} \approx 0.1$ nm $\gg \delta_p$.

As is evident from Eq. (6), at high powers the contribution of entangled pairs to the overall SFG rate becomes negligible. All previous experiments that involved SFG with down-converted light [17,18] were performed at extremely high powers ($\Phi \gg \Phi_{max}$). The observation and the temporal shaping of the coherent SFG process [denoted by the second term in Eq. (6)] in these experiments are a direct result of the fact that $\delta_{UC}$ was orders of magnitude smaller than the bandwidth of the light, thus violating the condition of Eq. (5). In such a case, many spectral combinations of the low-frequency fields contribute to the same narrowband up-converted field, and the resulting quantum interference can be controlled by manipulating the spectral phase of the incoming light, be it high-power down-converted light or shaped pulses. The effects observed in [17,18] do not reflect coincidences (nor the second-order correlation function of the light) and were demonstrated with shaped classical pulses as well [19]; had these experiments been performed with extremely ultrafast detectors (or with SFG with a large $\delta_{UC}$), the coincidence rate would have registered mostly the regular bunching expected from thermally distributed light.

The complementary aspect of the time and energy entanglement discussed earlier is the fact that the coherence time of the photon pairs ($\delta_p^{-1}$) is much larger than the individual photons' coherence time ($\Delta_{DC}^{-1}$), allowing two-photon interference even when the delay between the possible paths is too large to enable one-photon interference [1,2,20–23]. To observe such an interference we placed a Mach-Zehnder interferometer in the entangled-photon path after the last prism (see Fig. 1). The interferometer was constructed by rotating the polarization of the entangled photons to 45°, utilizing the SLM to induce retardation between the vertical and horizontal polarizations, and then rotating the polarization 45° back. In this configuration, the horizontal and vertical output polarizations are essentially the two output ports of the interferometer. In order to introduce retardations which were beyond the range of our SLM, we placed a 1 mm birefringent calcite crystal immediately after the SLM, thus adding a constant retardation of about 163 $\mu$m.

The dependence of the coincidence rate on the delay $\tau$ between the interferometer arms is proportional to $G^{(2)}(\tau)$ and can be represented by

$$R(\tau) \propto \left| \int g(\omega)\{\cos(\omega_o \tau) + \cos[(\omega - \omega_o)\tau]\} d\omega \right|^2, \quad (8)$$

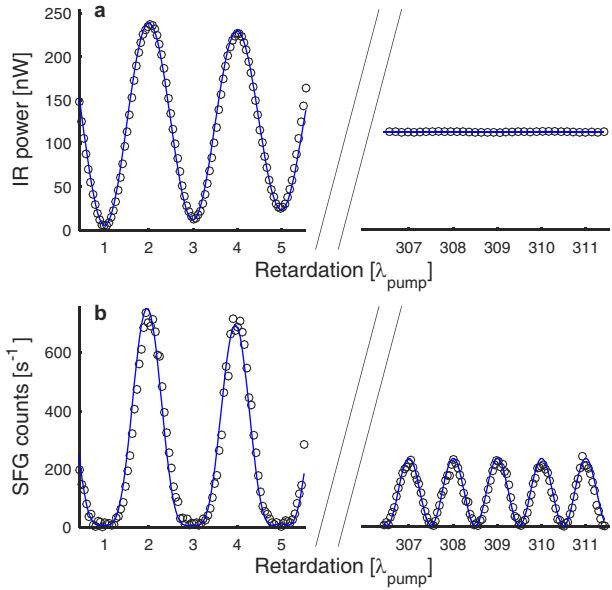

FIG. 3 (color online). Two-photon interference of time and energy entangled photons in a Mach-Zehnder interferometer (see Fig. 1). Interference oscillations of the IR power (a) and the SFG rate (b) as a function of the relative retardation between the interferometer arms. The experimental SFG counts (circles) are compared to the interference oscillations of the two-photon wave function, calculated according to Eq. (8) (line).

where the symmetry of $g(\omega)$ about $\omega_o = \omega_p/2$ was taken into account. Considering Eq. (8), we see that at small retardations the SFG rate is proportional to $|[\cos(\omega_o \tau) + 1]|^2$; i.e., it is quadratically proportional to the IR intensity oscillations, as classically expected. For retardations which are beyond the single-photon coherence length, the IR oscillations are washed out, and so is the second term in the integrand of Eq. (8), for the same reasons. Thus, the SFG rate becomes proportional to $|\cos(\omega_o \tau)|^2$, which oscillates with periodicity of $2\omega_o = \omega_p$. The two-photon interference oscillations maintain their high visibility due to the long coherence length of the entangled pairs, which enables interference between the events at which a photon pair traveled unseparated along one interferometer arm, or the other. However, whenever such oscillations were detected with photoelectric coincidences [21,23], the visibility dropped below 50% when the relative delay ranged between the coherence time of the photons (~100 fs) to the 4 orders of magnitude longer temporal resolution of the coincidence circuits (~1 ns). In contrast, the temporal resolution of the SFG process is equal to the coherence time of the photons since $\Delta_{LF} = \Delta_{DC}$. Thus, it performs as a "background-free" coincidence detector, inherently rejecting noncoinciding photons. As can be seen in Fig. 3, the SFG counts follow the interference oscillations of the two-photon wave function throughout the entire range of delays, demonstrating visibility of 94% ± 4% at a relative delay of ~550 fs, where the single-photon interference completely dies out. Note that 50% is the classical limit in the absence of first-order interference [21,23,24].

To conclude, we demonstrated how pulse-shaping techniques and SFG can be exploited to precisely control and measure the temporal properties of entangled-photon pairs. Although we emphasized the ability of SFG to directly reflect the shape of the two-photon wave function, spectral-phase manipulations can be observed and utilized in schemes based on coincidence measurements as well [25]. We believe both tools have potential applications in quantum measurement and quantum information science.

# Chapter 8
# Conclusions and Future Outlook

Broadband down converted light has very unique properties that are appealing both classically and quantum mechanically, to many applications. It uniquely combines incoherence in one photon with coherence in two photons, i.e. – it is a broadband noise field that is constrained to have only one quadrature in the complex plane. These properties are equivalent to those of ultrashort pulses in time and to narrowband CW lasers in frequency, thus making this light attractive to a number of applications. In this thesis, the basic properties of broadband down-converted light were demonstrated and manipulated, several applications were developed and a novel efficient source was designed and evaluated.

In the future, many paths can be taken, most of them unknown yet, to further pursue the research. First, a detailed experimental characterization and development of the OPO source is needed, both classically and quantum mechanically. Specifically, the current OPO cavity configuration can only be considered as a proof of principle. Further research of other configurations that will emit a stable broad spectrum is needed for viable applications. Such configurations should include engineered two-photon loss media with the desired dispersion properties. In addition, considerable reduction of the intra-cavity loss is necessary in order to be able to utilize the quantum mechanical potential of the OPO for emitting broadband highly squeezed light. An exciting direction for loss reduction is to develop fiber configurations of an OPO that may also be based on a different two-photon gain process (four waves mixing instead of three waves mixing, for example). Fiber based configurations are also very attractive for applications in optical communication.

Second, classical and quantum applications need more attention. A viable optical spread spectrum communication configuration must still be investigated. The same is true also for viable optical lithography and optical coherence tomography applications. In addition, research should be conducted on the quantum mechanical squeezing properties of high-power broadband down-conversion, which hold promise for exciting applications, such as measurement of optical phases at the Heisenberg limit of resolution (sub-shot noise) [57-59] or suppression of spontaneous atomic decay [60]. The squeezing properties are highly sensitive to loss, so a low loss OPO configuration is needed.

Finally, the quantum mechanical properties of two-photon processes, such as two-photon absorption and sum-frequency generation require more attention. Specifically, the possibility to use these processes as physical coincidence detectors is of great interest for two reasons. First, these processes allow simultaneous detection of both the time-difference and the energy-sum of the photons, thus allowing considerable rejection of noise (sum-frequency generation allows also measurement of the phase of photon-pairs). Second, although the two-photon detection efficiency of two-photon processes is very low, when the detector "clicks", a coincidence is guaranteed with femtosecond resolution (allowing Heisenberg limited phase measurements). Thus, the possibility to enhance the efficiency of such processes, either with resonant cavities or just by using media with stronger non-linearities, is of great interest for future research.

# Appendix A

# List of additional publications

1. B. Dayan, A. Pe'er, A. A. Friesem, and Y. Silberberg, "Coherent control with broadband squeezed vacuum", quant-ph/ p. 0302038 (2003), http://xxx.lanl.gov/abs/quant-ph/0302038
2. N. Dudovich, T. Polack, A. Pe'er and Y. Silberberg, " Simple Route to Strong-Field Coherent Control", Phys. Rev. Lett **94**, 083002 (2005).

# Appendix B

# Statement regarding my contribution to the papers included in this thesis

1. "Two Photon Absorption and Coherent Control with Broadband Down-Converted Light", B. Dayan, A. Pe'er, A. A. Friesem, Y. Silberberg, Physical Review Letters 93, 023005 (2004). One of the two main, equally contributing authors (together with Barak Dayan).

2. "Optical Code-Division Multiple Access Using Broad-Band Parametrically Generated Light", A. Pe'er, B. Dayan, Y. Silberberg, A. A. Friesem, Journal of Lightwave Technology 22, 1463-1471 (2004). Main contributor

3. "Quantum lithography by coherent control of classical light pulses", A. Pe'er, B. Dayan, M. Vucelja, Y. Silberberg, A. A. Friesem, Optics Express 12, 6600-6605 (2004). Main contributor

4. "An Intense Continuous Source of Broadband Down-Converted Light", A. Pe'er, Y. Silberberg, B. Dayan, A. A. Friesem, Submitted to Physical Review Letters (2005). Main contributor

5. "Nonlinear Interactions with an Ultrahigh Flux of Broadband Entangled Photons", B. Dayan, A. Pe'er, A. A. Friesem, Y. Silberberg, Physical Review Letters 94, 043602 (2005). One of the two main, equally contributing authors (together with Barak Dayan).

6. "Temporal Shaping of Entangled Photons", A. Pe'er, B. Dayan, A. A. Friesem, Y. Silberberg, Physical Review Letters 94, 073601 (2005). One of the two main, equally contributing authors (together with Barak Dayan).

# קורלציה ספקטרלית באור רחב סרט – תכונות, מקורות ושימושים

מוגש ע״י

## אבי פאר

בהנחיית פרופסור אשר פריזם

למועצה המדעית של מכון ויצמן למדע כחלק מהדרישות לקבלת תואר

דוקטור לפילוסופיה

מכון ויצמן למדע, רחובות ישראל

סיון התשס״ה (מאי 2005)

# תקציר

עבודה זו חוקרת באופן ניסיוני ותאורטי את תכונות הקוהרנטיות המיוחדות של אור "מומר מטה" (down converted light) הן במסגרת התאוריה הקלאסית והן הקוואנטית. למרות שהתכונות החד-פוטוניות של אור זה זהות לאלה של רעש לא קוהרנטי, יש לו תכונות קוהרנטיות דו-פוטונית ייחודיות. לכן, לצורך תהליכים דו-פוטוניים אור "מומר מטה" הוא שקול לפולסים קוהרנטיים אולטרא קצרים רחבי סרט מחד גיסא וללייזרים קוהרנטיים רציפים צרי סרט מאידך גיסא. שקילות זו מוכחת כאן תאורטית ומודגמת ניסיונית במשטר הקלאסי של עוצמות אור גבוהות. שני שימושים לאור זה מתוארים בפרוטרוט. האחד מנצל את תכונות הקוהרנטיות הדו-פוטונית הייחודיות לצורך מריחת תדר של תקשורת אופטית. האחר מדגים את היכולת לעבור בפקטור $N$ את גבול העקיפה הקלאסי על רזולוציה של מערכות ליטוגרפיה אופטית ע"י שימוש באור קלאסי בעל קוהרנטיות $N$-פוטונית, בין אם פולסי או "מומר מטה". חלק מפתח במחקר זה הוא המימוש של מקור אור הפולט אור "מומר מטה" רחב סרט בעצמה גבוהה והמבוסס על מהוד אופטי. תנודה רחבת סרט מובטחת במהוד ע"י מכניזם הדומה מאד ל"נעילת מודים" פסיבית במהודי לייזר הפולטים פולסים קצרים ולכן מכונה על ידנו "נעילת מודים בזוגות", מה שמרחיב את השקילות של אור "מומר מטה" לפולסים קצרים גם לעבר מקורות האור. לבסוף, התכונות הקוונטיות של אור "מומר מטה" רחב סרט נחקרות באופן ניסיוני ע"י ביצוע תהליך דו-פוטוני (יצירת סכום תדר) בעוצמות אור נמוכות מאוד של זוגות פוטונים בודדים מצומדים קוונטית. האופי הלא-קלאסי של האור מודגם ע"י תלות עצמה לינארית של התהליך הלא לינארי. תוך שימוש באינטראקציה לא לינארית זו בין פוטונים בודדים מודגם עיצוב בזמן של פונקציית הגל הדו-פוטונית ברזולוציה של פמטו-שניות.